\documentclass[11pt,usletter]{article}
\usepackage{jheppub}
\pdfoutput=1
\usepackage{graphicx}
\usepackage{hyperref}
\usepackage{color}

\def\ba{\begin{align}}
\def\ea{\end{align}}
\def\be{\begin{equation}}
\def\ee{\end{equation}}
\def\bea{\begin{eqnarray}}
\def\eea{\end{eqnarray}}

\begin{document}

\title{Supersymmetric SYK model and random matrix theory}

\date{\today}

\author[a]{Tianlin Li}
\author[b]{Junyu Liu}
\author[c]{Yuan Xin}
\author[d]{Yehao Zhou}

\affiliation[a]{
  Department of Physics and Astronomy, University of
  Nebraska,\\
  1400 R Street, Lincoln, Nebraska 68588, USA} 
\affiliation[b]{
  Walter Burke Institute for Theoretical Physics, California Institute of
  Technology,\\
  1200 east California Boulevard, Pasadena, California 91125, USA}
\affiliation[c]{
  Department of Physics, Boston University, \\
  590 Commonwealth Avenue, Boston, MA 02215, USA} 
\affiliation[d]{
  Perimeter Institute for Theoretical Physics, \\
  31 Caroline Street North, Waterloo, Ontario N2L 2Y5, Canada}

\emailAdd{tli11@unl.edu}
\emailAdd{jliu2@caltech.edu}
\emailAdd{yuan2015@bu.edu}
\emailAdd{yzhou3@perimeterinstitute.ca}

\abstract{In this paper, we investigate the effect of supersymmetry on the symmetry classification of random matrix theory ensembles. We mainly consider the random matrix behaviors in the $\mathcal{N}=1$ supersymmetric generalization of the Sachdev-Ye-Kitaev (SYK) model, a toy model for the two-dimensional quantum black hole with supersymmetric constraint. Some analytical arguments and numerical results are given to show that the statistics of the supersymmetric SYK model could be interpreted as random matrix theory ensembles, with a different eight-fold classification from the original SYK model and some new features. The time-dependent evolution of the spectral form factor is also investigated, where predictions from random matrix theory are governing the late time behavior of the chaotic Hamiltonian with supersymmetry.}


\maketitle
\flushbottom
\section{Introduction}
Physical systems with some stochastic or chaotic properties have some randomness in the setup of the fundamental Hamiltonian, which could be effectively simulated in the context of random matrix theory. When choosing an ensemble from random matrix theory for a chaotic Hamiltonian, we often need to consider the symmetries in the dynamics of the related physical system. The choice of standard matrix ensembles from symmetries historically comes from the invention of Dyson \cite{testbook}, which is called the three-fold way when classifying the Gaussian Unitary Ensemble (GUE), the Gaussian Orthogonal Ensemble (GOE), and the Gaussian Symplectic Ensemble (GSE). For more general symmetry discussion of interaction systems, the Altland-Zirnbauer theory gives a complete description as a ten-fold classification \cite{Zirnbauer1996,AZ1997}. In practical usage, one of the most celebrated works would be the classification of interaction inside topological insulators and topological phases in a ten-fold way \cite{ludwig,ki}.
\\
\\
In the recent study, the rising interests of studies on the Sachdev-Ye-Kitaev (SYK) model give another profound application in the random matrix theory classification. SYK model \cite{kitaev,Sachdev:1992fk} is a microscopic quantum Hamiltonian with random Gaussian non-local couplings among Majonara fermions. As is maximally chaotic and nearly conformal, this model could be treated as a holographic dual of the quantum black hole with $\text{AdS}_2$ horizon through the (near) AdS/CFT correspondence \cite{Almheiri:2014cka,Cvetic:2016eiv,Fu:2016yrv,Polchinski:2016xgd,Jevicki:2016bwu,Maldacena:2016hyu,Jensen:2016pah,Jevicki:2016ito,Bagrets:2016cdf,Maldacena:2016upp}. In recent research, people have also discussed several generalizations of the SYK model
\cite{Gu:2016oyy,Gross:2016kjj,Berkooz:2016cvq,Fu:2016vas}, such as higher dimensional generalizations and supersymmetric constraints. Some other related issues and similar models are discussed in
\cite{old,Sachdev2,Sannomiya:2016mnj,Sachdev:2010um,Garcia-Alvarez:2016wem,Hayden:2007cs,Anninos:2013nra,Sachdev:2015efa,Perlmutter:2016pkf,Anninos:2016szt,Danshita:2016xbo,Roberts:2016hpo,Betzios:2016yaq,Witten:2016iux,Patel:2016wdy,Klebanov:2016xxf,Blake:2016jnn,Nishinaka:2016nxg,Davison:2016ngz,Anninos:2016klf,Liu:2016rdi,Magan:2016ehs,Peng:2016mxj,Krishnan:2016bvg,Turiaci:2017zwd,Ferrari:2017ryl,Garcia-Garcia:2017pzl,Bi:2017yvx,Ho:2017nyc}. In the recent discussions, people have discovered that the SYK Hamiltonian has clear correspondence with the categories of the three-fold standard Dyson ensembles, unitary, orthogonal and symplectic ensembles, in the random matrix theory \cite{You:2016ldz,Garcia-Garcia:2016mno,Dyer:2016pou,Cotler:2016fpe}. In the recent work, \cite{Dyer:2016pou,Cotler:2016fpe}, it is understood that the time-dependent quantum dynamics of the temperature-dependent spectral form factor, namely, the combinations of partition functions with a special analytic continuation in SYK model, is computable in the late time by form factors in the random matrix theory with the same analytic continuation, as a probe of the discrete nature of the energy spectrum in a quantum black hole, and also a solid confirmation on the three-fold classification \cite{Cotler:2016fpe}.
\\
\\
In the route towards Dyson's classification, one only considers the set of simple unitary or anti-unitary operators as symmetries when commuting or anticommuting with the Hamiltonian. An interesting question would be, what is the influence of supersymmetry, the symmetry between fermions and bosons in the spectrum, in the classification of symmetry class?
\\
\\
As is illuminated by research in the past, supersymmetry \cite{Sohnius:1985qm} has several crucial influences in the study of disorder system and statistical physics \cite{Efetov:1997fw}, and could be emergent from condensed matter theory models \cite{Lee:2010fy}. Supersymmetry will enlarge the global symmetry group in the theory, has fruitful algebras and strong mathematical power used in several models in quantum mechanics and quantum field theory, and is extremely useful to simplify and clarify classical or quantized theories. In the recent study of the SYK model, the supersymmetric generalization for the original SYK has been discussed in detail in \cite{Fu:2016vas}, which shows several different behaviors through supersymmetric extensions. This model might give some implications in the quantum gravity structure of black hole in two dimensions in a supersymmetric theory, and also a related conjecture in  \cite{Cotler:2016fpe} for spectral form factor and correlation functions in super Yang-Mills theory.
\\
\\
In order to explore the supersymmetric constraints on the random matrix theory classification, in this paper, we will study the symmetry classification and random matrix behavior of the $\mathcal{N}=1$ supersymmetric extension of SYK model by Fu-Gaiotto-Maldacena-Sachdev's prescription \cite{Fu:2016vas}. The effect of supersymmetry in the symmetry classification could be summarized as the following. Supersymmetry will cause the Hamiltonian to show a quadratic expression. Namely, we could write $H$ as the square of $Q$. This condition will greatly change the distribution of the eigenvalues. From random matrix language \cite{statistical}, if $Q$ is a Gaussian random matrix, then $H$ should be in a Wishart-Laguerre random matrix, with the eigenvalue distribution changing from Wigner's semi-circle to the Marchenko-Pastur distribution. In another sense, the quadratic structure will fold the eigenvalues of $Q$ and cause a positivity condition for all eigenvalues. Namely, if $Q$ has the eigenvalue distribution that eigenvalues come in pair with positive and negative signs, the squaring $Q$ will cause larger degeneracies and a folded structure in eigenvalues of energy. Moreover, the coupling degree might be changed when considering $Q$ instead of $H$. For instance, in the $\mathcal{N}=1$ extended SYK model, $Q$ is a non-local three-point coupling, which is not even. This will change the previous classification in the Hamiltonian based on the representation of Clifford algebra from a mathematical point of view.
\\
\\
These aspects will be investigated in a clearer and more detailed way in the paper.
\\
\\
This paper will be organized as the following. In Section \ref{models}, we will review the model construction and thermodynamics of the SYK model and its supersymmetric extensions. In Section \ref{RMT}, we will discuss the random matrix classification for models, especially supersymmetric extensions of the SYK model. In Section \ref{data}, we will present our numerical confirmation for symmetry classifications from the exact diagonalization, including the computation of the density of states and spectral form factors. In Section \ref{conclu}, we will arrive at a conclusion and discuss the directions for future works. In the appendix, we will review some knowledge to make this paper self-contained, including basics on Altland-Zirnbauer theory and a calculation on the random matrix theory measure.
\section{Introduction on models}\label{models}
In this paper, we will mostly focus on SYK models and their extensions. Thus before the main investigation, we will provide a simple introduction to the necessary knowledge of related models to be self-contained.
\subsection{The SYK model}
In this part, we will simply review the SYK model, mainly following \cite{Maldacena:2016hyu}. The SYK model is a microscopic model with some properties of the quantum black hole. The Hamiltonian\footnotemark is given by
\begin{align}
H=\sum\limits_{i<j<k<l}{{{J}_{ijkl}}{{\psi }^{i}}{{\psi }^{j}}{{\psi }^{k}}{{\psi }^{l}}}
\end{align}
where $\psi^i$ are Majorana fermions, and they are coupled by the four-point random coupling with Gaussian distribution
\begin{align}
\left\langle {{J}_{ijkl}} \right\rangle =0~~~~~~\left\langle J_{ijkl}^{2} \right\rangle =\frac{6J_{\text{SYK}}^{2}}{{{N}^{3}}}=\frac{12\mathcal{J}_{\text{SYK}}^{2}}{{{N}^{3}}}
\end{align}
where $J_\text{SYK}$ and $\mathcal{J}_\text{SYK}$ are positive constants, and $J_\text{SYK}=\sqrt{2}\mathcal{J}_\text{SYK}$. The large $N$ partition function is given by
\begin{align}
 Z(\beta )\sim\exp (-\beta {{E}_{0}}+N{{s}_{0}}+\frac{cN}{2\beta })
\end{align}
where $E_0$ is the total ground state energy proportional to $N$ and it is roughly $E_0=-0.04 N$ \cite{Cotler:2016fpe}. $s_0$ is the ground state entropy contributed from one fermion, and one can estimate it theoretically \cite{Maldacena:2016hyu},
\begin{align}
{{s}_{0}}=\frac{G}{2\pi }+\frac{\log 2}{8}=0.2324
\end{align}
where $G$ is the Catalan number. $c$ is the specific heat, which could be computed by
\begin{align}
c=\frac{4{{\pi }^{2}}{{\alpha }_{S}}}{\mathcal{J}_\text{SYK}}=\frac{0.3959}{J_\text{SYK}}
\end{align}
and $\alpha_S=0.0071$ is a positive constant. This contribution $c/\beta$ is from the Schwarzian, the quantum fluctuation near the saddle point of the effective action in the SYK model. The Schwarzian partition function is
\begin{align}
{{Z}_\text{Sch}}(\beta)\sim\int{\mathcal{D}\tau (u)}\exp \left( -\frac{\pi N{{\alpha }_{S}}}{\beta \mathcal{J}_\text{SYK}}\int_{0}^{2\pi }{du\left( \frac{\tau '{{'}^{2}}}{\tau {{'}^{2}}}-\tau {{'}^{2}} \right)} \right)
\end{align}
where the path integral is taken for all possible reparametrizations $\tau(u)$ of the thermal circle in different equivalent classes of the $\text{SL}(2,\mathbb{R})$ symmetry. The Schwarzian corresponds to the broken reparametrization symmetry of the SYK model. One can compute the one-loop correction from the soft mode of the broken symmetry,
\begin{align}
{{Z}_{\text{Sch}}}(\beta )\sim\frac{1}{{{(\beta J_\text{SYK})}^{3/2}}}\exp \left( \frac{cN}{2\beta } \right)
\end{align}
As a result, one can consider the correction from the soft mode if we consider an external one-loop factor $(\beta J_\text{SYK})^{-3/2}$. The density of states could also be predicted by the contour integral of the partition function as
\begin{align}
\rho (E)\sim\exp (N{{s}_{0}}+\sqrt{2cN(E-{{E}_{0}})})
\end{align}
\footnotetext{One could also generalize the SYK model to general $q$-point non-local interactions where $q$ are even numbers larger than four. The Hamiltonian should be,
\begin{align}
H=i^{q/2}\sum\limits_{{{i}_{1}}<{{i}_{2}}<\ldots <{{i}_{q}}}{{{J}_{{{i}_{1}}{{i}_{2}}\ldots {{i}_{q}}}}{{\psi }^{{{i}_{1}}}}{{\psi }^{{{i}_{2}}}}\ldots {{\psi }^{{{i}_{q}}}}}
\end{align}
where
\begin{align}
\left\langle {{J}_{{{i}_{1}}{{i}_{2}}\ldots {{i}_{q}}}} \right\rangle =0~~~~~~\left\langle J_{{{i}_{1}}{{i}_{2}}\ldots {{i}_{q}}}^{2} \right\rangle =\frac{{J_\text{SYK}^2}(q-1)!}{{{N}^{q-1}}}=\frac{{{2}^{q-1}}}{q}\frac{{\mathcal{J}_{\text{SYK}}^{2}}(q-1)!}{{{N}^{q-1}}}
\end{align}
Sometimes we will discuss general $q$s in this paper but we will mainly focus on the $q=4$ case.
}
\subsection{$\mathcal{N}=1$ supersymmetric extension}
Following \cite{Fu:2016vas}, in the supersymmetric extension of the SYK model, firstly we define the supercharge\footnotemark
\begin{align}
Q=i\sum\limits_{i<j<k}{{{C}_{ijk}}{{\psi }^{i}}{{\psi }^{j}}{{\psi }^{k}}}
\end{align}
for Majonara fermions $\psi^i$. $C_{ijk}$ is a random tensor with the Gaussian distribution as the coupling,
\begin{align}
\left\langle {{C}_{ijk}} \right\rangle =0 ~~~~~~\left\langle C_{ijk}^{2} \right\rangle =\frac{2J_{\mathcal{N}=1}}{{{N}^{2}}}
\end{align}
where $J_{\mathcal{N}=1}$ is also a constant with mass dimension one. The square of the supercharge will give the Hamiltonian of the model
\begin{align}\label{susyHamiltonian}
H={{E}_{c}}+\sum\limits_{i<j<k<l}{{{J}_{ijkl}}{{\psi }^{i}}{{\psi }^{j}}{{\psi }^{k}}{{\psi }^{l}}}
\end{align}
where
\begin{align}\label{susyenergy}
  & {{E}_{c}}=\frac{1}{8}\sum\limits_{i<j<k}{C_{ijk}^{2}} ~~~~~~{{J}_{ijkl}}=-\frac{1}{8}\sum\limits_{a}{{{C}_{a[ij}}{{C}_{kl]a}}}
\end{align}
where $[\cdots]$ is the summation of all possible antisymmetric permutations. Besides the shifted constant $E_c$, the distribution of $J_{ijkl}$ is different from the original SYK model because it is not a free variable of Gaussian distribution, which changes the large $N$ behavior of this model. In the large $N$ limit, the model has an unbroken supersymmetry with a bosonic superpartner $b^i$. The Lagrangian of this model is given by
\begin{align}
L=\sum\limits_{i}{\left( \frac{1}{2}{{\psi }^{i}}{{\partial }_{\tau }}{{\psi }^{i}}-\frac{1}{2}{{b}^{i}}{{b}^{i}}+i\sum\limits_{j<k}{{{C}_{ijk}}{{b}^{i}}{{\psi }^{j}}{{\psi }^{k}}} \right)}
\end{align}
In this model, the Schwarzian is different from the original SYK model. We also have the expansion for the large $N$ partition function
\begin{align}
 Z(\beta )\sim\exp (-\beta {{E}_{0}}+N{{s}_{0}}+\frac{cN}{2\beta })
\end{align}
But the results of $E_0$ and $s_0$ are different (while the specific heat is the same for these two models). In the large $N$ limit, the supersymmetry is preserved. Thus we have the ground state energy $E_0=0$. The zero-temperature entropy is given by
\begin{align}
{{s}_{0}}=\frac{1}{2}\log (2\cos \frac{\pi }{6})=\frac{1}{4}\log 3=0.275
\end{align}
Moreover, the one-loop correction from Schwarzian action is different. As a result of supersymmetry constraint, the one-loop factor is $(\beta J_{\mathcal{N}=1})^{-1/2}$
\begin{align}
{{Z}_\text{Sch}}(\beta )\sim\frac{1}{{{(\beta J_{\mathcal{N}=1})}^{1/2}}}{{e}^{N{{s}_{0}}+cN/2\beta }}
\end{align}
which predicts a different behavior for the density of states
\begin{align}
\rho (E)\sim\frac{1}{({EJ_{\mathcal{N}=1}})^{1/2}}{{e}^{N{{s}_{0}}+2c N E }}
\end{align}
\footnotetext{For the generic positive integer $\hat{q}$ we can also define the $\mathcal{N}=1$ supersymmetric extension with non-local interaction of $2\hat{q}-2$ fermions. The supercharge should be
\begin{align}
Q={{i}^{\frac{\hat{q}-1}{2}}}\sum\limits_{{{i}_{1}}<{{i}_{2}}<\ldots <{{i}_{{\hat{q}}}}}{{{C}_{{{i}_{1}}{{i}_{2}}\ldots {{i}_{{\hat{q}}}}}}{{\psi }^{{{i}_{1}}}}{{\psi }^{{{i}_{2}}}}\ldots {{\psi }^{{{i}_{{\hat{q}}}}}}}
\end{align}
where
\begin{align}
\left\langle {{C}_{{{i}_{1}}{{i}_{2}}\ldots {{i}_{{\hat{q}}}}}} \right\rangle =0~~~~~~\left\langle C_{{{i}_{1}}{{i}_{2}}\ldots {{i}_{{\hat{q}}}}}^{2} \right\rangle =\frac{(\hat{q}-1)!J_{\mathcal{N}=1}}{{{N}^{\hat{q}-1}}}=\frac{{{2}^{\hat{q}-2}}(\hat{q}-1)!\mathcal{J}_{\mathcal{N}=1}}{q{{N}^{\hat{q}-1}}}
\end{align}
And $\hat{q}=3$ will recover the case in the main text.
}
\section{Random matrix classification}\label{RMT}
It is established that the SYK model is classified by random matrix theory in that the random interacting SYK Hamiltonian fall into one of the three standard Dyson ensembles in the eight-fold way \cite{You:2016ldz,Garcia-Garcia:2016mno,Dyer:2016pou,Cotler:2016fpe}. It is natural to believe that the supersymmetric extension can also be described by random matrix theory. To sharpen the argument, we derive the exact correspondence between each SYK Hamiltonian and some random matrix ensembles; in other words, the eight-fold rule for the supersymmetric case. A priori, the supersymmetric SYK Hamiltonian, should lead to a different random matrix theory description than the original case. Superficially, the original SYK theory and its supersymmetric cousin are different have two major differences, which have also been mentioned in the previous discussions.
\begin{itemize}
  \item The degeneracy of the two Hamiltonian matrices is different. The degeneracy of the supersymmetric SYK model is also investigated by \cite{Fu:2016vas}, which we derive again using some different discussion in Section~\ref{sec:RMTH}. The degeneracy space is enlarged by supersymmetry. Generally, the energy level distribution of random matrices is sensitive to the degeneracy and is thus sensitive to the supersymmetric extension.
  \item Another difference is the apparent positive semidefiniteness of the Hamiltonian being the square of the supercharge. We will see later that the positive constraint leads to a new eigenvalue distribution different from those of Gaussian ensembles.
\end{itemize}
Symmetry analysis is crucial in classifying the random matrix statistics of Hamiltonian matrices. \cite{You:2016ldz,Cotler:2016fpe} argue that the particle-hole symmetry operator determines the class of random matrix theory statistics. The random matrix classification dictionary is determined by the degeneracy and the special relations required by having the symmetry. The systematic method of random matrix classification is established as the Atland-Zirnbauer theory \cite{Zirnbauer1996,AZ1997}, reviewed in appendix \ref{AZ}. The anti-unitary operators play a central role in the classifications. The Atland Zirnbauer also applies to extended ensembles different from the three standard Dyson ensembles, which we find useful in classifying the supersymmetric SYK theory. In Section \ref{sec:RMTSYK}, we derive the eight-fold way classification of the original SYK Hamiltonian again using the Atland-Zirnbauer theory and find the matrix representations of Hamiltonian in each mod-eight sector unambiguously. We notice that the matrix representation of Hamiltonian takes block diagonal form, with each block being a random matrix from a certain ensemble. This block diagonal form is also found by \cite{You:2016ldz} in a different version.
\\
\\
Naively one would apply the same argument to the supersymmetric Hamiltonian, since it also enjoys the particle-hole symmetry. But this is not the full picture. First, one needs to take into account of Hamiltonian being the square of the supercharge and is thus not totally random. In Section~\ref{sec:RMTQ}, we argue that the supercharge $Q$ has a random matrix description that falls into one of the extended ensembles. Using the Atland-Zirnbauer theory on $Q$, we obtain its matrix representation in block diagonal form and use it to determine the matrix representation of the Hamiltonian in Section~\ref{sec:RMTH}. Second, in order to obtain the correct classification, one needs to consider the full set of symmetry operators. Apparently, particle-hole is not enough since supersymmetry enlarges the SYK degeneracy space. We argue that the Witten index operator, $(-1)^F$, is crucial in the symmetry analysis of any system with supersymmetry. Incorporating $(-1)^F$ we obtain the full set of symmetry operators. Finally, the squaring operation will change the properties of the random matrix theory distribution of the supercharge $Q$, from Gaussian to Wishart-Laguerre. The quantum mechanics and statistics in supersymmetric SYK models, based on the main investigation in this paper, might be a non-trivial and compelling example of the supersymmetric symmetry class.


\subsection{SYK}\label{sec:RMTSYK}
Now we apply the Altland-Zirnbauer classification theory (see appendix \ref{AZ} for some necessary knowledges) to the
original SYK model \cite{You:2016ldz,Garcia-Garcia:2016mno,Dyer:2016pou,Cotler:2016fpe}. This is accomplished by finding the symmetry of the theory (and has already been discussed in other works, see \cite{You:2016ldz,Cotler:2016fpe}). First, one can change the Majonara fermion operators to creation annihilation operators $c^\alpha$ and $\bar {c}^\alpha$ by
\begin{align} {{\psi
}^{2\alpha}}=\frac{{{c}^{\alpha}}+{{{\bar{c}}}^{\alpha}}}{\sqrt{2}}~~~~~~{{\psi
}^{2\alpha-1}}=\frac{i({{c}^{\alpha}}-{{{\bar{c}}}^{\alpha}})}{\sqrt{2}}
\end{align} 
where $\alpha = 1,2\cdots,N_d=N/2$. The fermionic number operator $F=\sum_\alpha\bar{c}^\alpha c^\alpha$ divides the total Hilbert space into two different charge parities. One can define the particle-hole operator
\begin{align}
P=K\prod\limits_{\alpha=1}^{N_d}{({{c}^{\alpha}}+{{\bar{c}}}^{\alpha})}
\end{align} 
where $K$ is the complex conjugate operator ($c^{\alpha}$ and $\bar{c}^{\alpha}$ are real). The operation of $P$ on fermionic operators is given by
\begin{align} P{{c}^{\alpha }}P=\eta {{c}^{\alpha
}}~~~~~~P{{{\bar{c}}}^{\alpha }}P=\eta {{{\bar{c}}}^{\alpha
}}~~~~~~P{{\psi }^{i}}P=\eta {{\psi }^{i}}
\end{align} 
where
\begin{align} \eta ={{(-1)}^{[3{{N}_{d}}/2-1]}}
\end{align} 
From these commutation relations we can show that
\begin{align} [H,P]=0
\end{align} 
To compare with the Altland-Zirnbauer classification, we need to know the square of $P$, and this is done by direct calculation
\begin{align}
{{P}^{2}}={{(-1)}^{[{{N}_{d}}/2]}}=\left\{ \begin{matrix} +1 & N\bmod
8=0 \\ +1 & N\bmod 8=2 \\ -1 & N\bmod 8=4 \\ -1 & N\bmod 8=6 \\
\end{matrix} \right.
\end{align} 
Now we discover that $P$ can be treated as a $T_+$ operator, and it completely determines the class of the Hamiltonian. Before we list the result, it should be mentioned that the degeneracy of Hamiltonian can be seen from the properties of $P$:
\begin{itemize}
\item $N\text{ mod }8=2\text{ or }6$:\\
  The symmetry $P$ exchanges the parity sector of a state, so there is a two-fold degeneracy. However, there are no further symmetries caused by $P$ in each block. Thus, it is given as a combination of two GUEs, where two copies of GUEs degenerate.
\item $N\text{ mod }8=4$: \\
  The symmetry $P$ is a parity-invariant mapping and $P^2=-1$, so there is a two-fold degeneracy. There are no further independent symmetries. From the Altland-Zirnbauer theory, we know that in each parity block, there is a GSE matrix. Also, where two copies of GSEs are independent.
\item $N\text{ mod }8=0$: \\
  The symmetry $P$ is a parity-invariant mapping and $P^2=1$. There are no further symmetries, so the degeneracy is one. From the Altland-Zirnbauer theory, we know that in each parity block, there is a GOE matrix. Also, two copies of GOEs are independent.
\end{itemize}
We summarize those observations in the following table as a summary of SYK model,
\begin{center}
\begin{tabular}{ c | c | c | c | c | c   }
$N \bmod 8$ & Deg. & RMT & Block & Type & Level stat.\\
\hline
0 & 1 & $\text{GOE}$ & $\left( \begin{matrix}
   A & 0  \\
   0 & B  \\
\end{matrix} \right)\text{ }A,B \text{ real symmetric}$
&$\mathbb{R}$& GOE\\
2 & 2 & $\text{GUE}$ &$\left( \begin{matrix}
   A & 0  \\
   0 & \bar{A}  \\
\end{matrix} \right)\text{  }A \text{ Hermitian}$
&$\mathbb{C}$& GUE\\
4 & 2 & $\text{GSE}$ &$\left( \begin{matrix}
   A & 0  \\
   0 & B  \\
 \end{matrix} \right)\text{  }A,B \text{ Hermitian quaternion}$ &$\mathbb{H}$& GSE\\
  6 & 2 & $\text{GUE}$ &$\left( \begin{matrix}
      A & 0  \\
      0 & \bar{A}  \\
\end{matrix} \right)\text{  }A\text{ Hermitian}$ &$\mathbb{C}$& GUE\\
\end{tabular}
\end{center}
where the level statistics means some typical numerical evidence of random matrix, for instance, the Wigner surmise, number variance, or $\Delta_3$ statistics, etc.. Although the SYK Hamiltonian can be decomposed as two different parity sectors, we can treat them as standard Dyson random matrix as a whole because these two sectors are either independent or degenerate (The only subtleties will be
investigating the level statistics when considering two independent sectors, where two mixed sectors will show a many-body localized phase statistics instead of chaotic phase statistics, which has been discussed originally in \cite{You:2016ldz}.) In the following, we will also numerically test the random matrix behavior, and based on the numerical testing range of $N$ we can summarize the following table for practical usage.
\begin{center}
\begin{tabular}{ c | c | c | c | c | c | c| c| c| c | c }
$N$ & 10 & 12 & 14 & 16 &18 & 20 & 22 & 24 & 26 & 28 \\
\hline
Ensemble & GUE& GSE& GUE & GOE& GUE& GSE & GUE & GOE & GUE & GSE\\
\end{tabular}
\end{center}
\subsection{$\mathcal{N}=1$ supersymmetric classification}
Supersymmetry algebra is a $\mathbb{Z}_2$-graded algebra, where states and operators are subdivided into two distinct parity sectors. In such an algebra, there may exist a Klein operator \cite{JunkerGeorg:2012} which anti-commutes with any operators with odd parity and commutes with any operators with even parity. The Klein operator of supersymmetry algebra is naturally the Witten index operator.
\\
\\
Witten index might play a role in the symmetric structure and block decomposition in the supersymmetric quantum mechanics. A simple example is \cite{JunkerGeorg:2012}, in $\mathcal{N}=2$ supersymmetry algebra, Define $W$ be the Witten operator. The Witten operator has eigenvalue $\pm 1$ and separates the Hilbert space into two parity sectors
\begin{equation}
  \mathcal{H} = \mathcal{H}^+ \oplus \mathcal{H}^-~.
\end{equation}
We can also define projection operators $P^\pm = 1/2(1\pm W)$. In the parity representation the operators take $2 \times 2$ block-diagonal form
\begin{equation}
  W = \left(
    \begin{array}{cc}
      1 & 0 \\
      0 & -1
    \end{array}
  \right)~, ~~~
  P^+ = \left(
    \begin{array}{cc}
      1 & 0 \\
      0 & 0
    \end{array}
  \right)~, ~~~
  P^- = \left(
    \begin{array}{cc}
      0 & 0 \\
      0 & 1
    \end{array}
  \right)~.
\end{equation}
Because of $Q^2=0$ and $\{Q,W\}=0$ the complex supercharges are necessarily of the form
\begin{equation}\label{eq:Z2gradingN2Generic}
  Q = \left(
    \begin{array}{cc}
      0 & A \\
      0 & 0
    \end{array}
  \right)~, ~~~
  Q^\dagger = \left(
    \begin{array}{cc}
      0 & 0 \\
      A^\dagger & 0
    \end{array}
  \right)~,
\end{equation}
which imply
\begin{equation}
  Q_1 = \frac{1}{\sqrt{2}}\left(
    \begin{array}{cc}
      0 & A \\
      A^\dagger & 0
    \end{array}
  \right)~, ~~~
  Q_2 = \frac{i}{\sqrt{2}}\left(
    \begin{array}{cc}
      0 & -A \\
      A^\dagger & 0
    \end{array}
  \right)~.
\end{equation}
In the above equation, $A$ takes $\mathcal{H}^- \rightarrow\mathcal{H}^+$ and its adjoint $A^\dagger$ takes $\mathcal{H}^+ \rightarrow\mathcal{H}^-$. The supersymmetric Hamiltonian becomes diagonal in this representation
\begin{equation}\label{eq:Z2gradedH}
  H=\left(
    \begin{array}{cc}
      AA^\dagger & 0 \\
      0 & A^\dagger A
    \end{array}
  \right)~.
\end{equation}
In this construction, the Hilbert space is divided by the Witten parity operator. The Hamiltonian is shown to take the block diagonal positive semidefinite form without even referring to the explicit construction of the Hamiltonian. It is remarkable that the above computation is very similar to our work from Section \ref{sec:RMTQ} to \ref{sec:RMTH}. 
\\
\\
The argument can also work in a reversive way. Hidden supersymmetry can be found in a bosonic system such as a Calogero-like model \cite{Calogero:1969xj}, a system of one-dimensional Harmonic oscillators with inverse square interactions and extensions. What makes supersymmetry manifest is the Klein operator. The model and its various extensions are studied in \cite{Brink:1993sz,Plyushchay:1994re,Plyushchay:1996ry,Bekaert:2005vh,Brink:1992xr}. A trivial simple Harmonic operator has algebra $ \left[ a^- , a^+ \right] = 1$. The algebra describes a bosonic system. $\mathbb{Z}_2$ grading is realized by introducing an operator $K = \cos (\pi a^+ a^-)$. The new operator anti-commutes with $a^-$ and $a^+$; thus is a Klein operator. Based on the Klein operator, one can construct the projection operators in both sectors and also the supercharge. In this way, the simple harmonic oscillator is ``promoted'' to have supersymmetry. A generalization to simple harmonic oscillator is the deformed Heisenberg algebra, $\left[ a^- , a^+ \right] = 1 + \nu K$. The corresponding system is an $\mathcal{N}=2$ supersymmetric extension of the 2-body Calogero. The model is also used in a considerably simplifying Calogero model.
\\
\\
The above observations strongly support the argument that supersymmetry will change the classification of symmetry class in quantum mechanical models. In the following work, we will show that the supersymmetric SYK model symmetry class can be explicitly constructed and change the classification of random matrix theory ensembles.
\subsubsection{Supercharges in $\mathcal{N}=1$ SYK}
\label{sec:RMTQ}
In the $\mathcal{N}=1$ supersymmetric model, it should be more
convenient to consider the spectrum of $Q$ instead of $H$, because $H$ is a square of $Q$. Although $Q$ is not a Hamiltonian, since we only
care about its matrix type, and the Altland-Zirnbauer theory is purely
mathematical, $Q$ can be treated as a Hamiltonian. Similar to the
original SYK model, we are concerned about the symmetry of the
theory. We notice that the Witten index $(-1)^F$ is
\begin{align}
  (-1)^F =(-2i)^{N_d}\prod _{i=1}^{N}\psi ^i= \prod
_{\alpha=1}^{N_d}(1-2\bar {c}^\alpha c^\alpha)
\end{align} 
which is the fermionic parity operator up to a sign $(-1)^{N_d}$. Witten index and particle-hole symmetry have the following commutation relation:
\begin{align} 
P(-1)^F=(-1)^{N_d} (-1)^F P
\end{align} 
Now we define a new operator, $R=P(-1)^F$. It has a compact form
\begin{align} 
R=K\prod _{\alpha=1}^{N_d} (c^\alpha-\bar {c}^\alpha)
\end{align} 
$R$ and $P$ are both anti-unitary symmetries of $Q$, with commutation relations:
\begin{center}
\begin{tabular}{ c | c | c } $N$ mod 8 & $P$ & $R$ \\ \hline 0 &
$[P,Q]=0$ & $\{R,Q\}=0$\\ 2 & $\{P,Q\}=0$ & $[R,Q]=0$\\ 4 & $[P,Q]=0$
& $\{R,Q\}=0$\\ 6 & $\{P,Q\}=0$ & $[R,Q]=0$\\
\end{tabular}
\end{center} and squares
\begin{align} {{P}^{2}}={{(-1)}^{[{{N}_{d}}/2]}}\text{,
}R^2=(-1)^{[N_d/2]+N_d}
\end{align} 
Thus, in different values of $N$, the two operators $P$
and $R$ behave differently and replace the role in $T_+$ and $T_-$ in
the Altland-Zirnbauer theory. Now we can list the classification for
the matrix ensemble of $\mathcal N=1$ supersymmetric SYK model,
\begin{center}
\begin{tabular}{ c | c | c | c | c | c} $N$ mod 8 & $T_+^2$ & $T_-^2$
& $\Lambda^2$ & Cartan Label & Type \\ \hline 0 & $P^2=1$ & $R^2=1$ &
1& BDI (chGOE) & $\mathbb{R}$ \\ 2 & $R^2=-1$ & $P^2=1$ & 1& DIII
(BdG) & $\mathbb{H}$ \\ 4 & $P^2=-1$ & $R^2=-1$ & 1& CII (chGSE) &
$\mathbb{H}$\\ 6 & $R^2=1$ & $P^2=-1$ & 1& CI (BdG) & $\mathbb{R}$ \\
\end{tabular}
\end{center} One can also write down the block representation of
$Q$. Notice that the basis of block decomposition is based on the
$\pm 1$ eigenspaces of anti-unitary operators, namely, it is
decomposed based on the parity.
\subsubsection{Hamiltonians in the $\mathcal{N}=1$ theory}\label{sec:RMTH}
Now we already obtain the random matrix type of the supercharge. Thus
the structure of the square of $Q$ could be considered case by case. Before that, we can notice one general property, that unlike the
GOE or GSE group in SYK, in the supersymmetric model, there is a supercharge $Q$ contains an odd number of Dirac fermions as a symmetry of $H$. Thus it always changes the parity. Thus the spectrum of $H$ is always decomposed to two degenerate blocks. Another general property is that the spectrum of $H$ is always positive because $Q$ is Hermitian and $H=Q^2>0$. Thus the random matrix class of
$\mathcal{N}=1$ will be some classes up to positivity constraint.
\begin{itemize}
\item $N=0 \bmod 8$:
In this case, $Q$ is a BDI (chGOE) matrix. Thus we can write down the block decomposition as
\begin{align}
Q=\left( \begin{matrix}
   0 & A  \\
   {{A}^{T}} & 0  \\
\end{matrix} \right)
\end{align}
where $A$ is a real matrix. Thus the Hamiltonian is obtained by
\begin{align}
H=\left( \begin{matrix}
   A{{A}^{T}} & 0  \\
   0 & {{A}^{T}}A  \\
\end{matrix} \right)
\end{align}
Since $ AA^{T}$ and $A^T A$ share the same eigenvalues ($\{R,Q\}=0$. Thus $R$ flips the sign of eigenvalues of $Q$, but after squaring these two eigenvalues with opposite signatures become the same), and there is no internal structure in $A$ (in this case $P$ is a symmetry of $Q$, $[P,Q]=0$, but $P^2=1$. Thus, $P$ cannot provide any further degeneracy), 
we obtain that $H$ has a two-fold degeneracy. Moreover, because $A A^T$ and $A^T A$ are both real positive-definite symmetric matrix without any further structure, it is nothing but the subset of GOE symmetry class with positivity condition. These two sectors will be exactly degenerate.
\item $N=4\text{ mod }8$: In this case $Q$ is a CII (chGSE)
matrix. Thus we can write down the block decomposition as
\begin{align} Q=\left( \begin{matrix} 0 & B \\ {{B}^{\dagger }} & 0 \\
\end{matrix} \right)
\end{align} 
where $B$ is a quaternion Hermitian matrix. Thus after
squaring we obtain
\begin{align} H=\left( \begin{matrix} B{{B}^{\dagger }} & 0 \\ 0 &
{{B}^{\dagger }}B \\
\end{matrix} \right)
\end{align}
Since $ B{{B}^{\dagger }}$ and ${{B}^{\dagger }}B$ share the same eigenvalues, and each block has a natural two-fold degeneracy by the property of quaternion (Physically it is because $\{R,Q\}=0$ thus $R$ flips the sign of eigenvalues of $Q$. But after squaring, these two eigenvalues with opposite signatures become the same. Moreover, in this case, $P$ is a symmetry of $Q$, $[P,Q]=0$, and $P^2=-1$), we get a four-fold degeneracy in the spectrum of $H$. Because $B B^\dagger$ and $B^\dagger B$ are quaternion Hermitian matrices when $B$ is quaternion Hermitian\footnotemark, $B B^\dagger=B^\dagger B$ are both quaternion Hermitian positive-definite matrix without any further structure. As a result, it is nothing but the subset of the GSE symmetry class with positivity condition.  These two sectors will be exactly degenerate.
\item $N=2\text{ mod }8$:
In this case, $Q$ is a DIII (BdG) matrix. Thus we can write down the block decomposition as
\begin{align}
Q=\left( \begin{matrix}
   0 & Y  \\
   -\bar{Y} & 0  \\
\end{matrix} \right)
\end{align}
where $Y$ is a complex, skew-symmetric matrix. Thus after squaring, we obtain
\begin{align}
H=\left( \begin{matrix}
   -Y\bar{Y} & 0  \\
   0 & -\bar{Y}Y  \\
\end{matrix} \right)
\end{align}
Firstly let us take a look at the degeneracy. Since $-Y\bar{Y}$ and $-\bar{Y}Y$ share the same eigenvalues, and each block has a natural two-fold degeneracy because in the skew-symmetric matrix the eigenvalues come in pair and after squaring pairs coincide (Physically, it is because $\{P,Q\}=0$. Thus $P$ flips the sign of eigenvalues of $Q$, but after squaring, these two eigenvalues with opposite signatures become the same. Also, in this case $R$ is a symmetry of $Q$, $[R,Q]=0$, and $R^2=-1$), we obtain a four-fold spectrum of $H$.
\\
\\
Now take the operator $Q$ as a whole, from the previous discussion, we may note that it is quaternion Hermitian because it could be easily verified that $Q\Omega=\Omega Q$ and $Q^\dagger=Q$. Thus $Q^2=H$ must be a quaternion Hermitian matrix (there is another way to see that, which is taking the block decomposition by another definition of quaternion Hermitian, squaring it and check the definition again). Moreover, $H$ has a two-fold degenerate parity decomposition. Thus, in each part, it is also a quaternion Hermitian matrix. Because in the total matrix, it is a subset of GSE symmetry class (with positivity constraint), in each degenerate parity sector, it is also in a subset of positive definite GSE symmetry class (one can see this by applying the total measure in the two different, degenerate part).

\item $N=6\text{ mod }8$:
In this case, $Q$ is a CI (BdG) matrix. Thus, we can write down the block decomposition as
\begin{align}
Q=\left( \begin{matrix}
   0 & Z  \\
   {\bar{Z}} & 0  \\
\end{matrix} \right)
\end{align}
where $Z$ is a complex symmetric matrix. Thus, after squaring, we obtain
\begin{align}
H=\left( \begin{matrix}
   Z\bar{Z} & 0  \\
   0 & \bar{Z}Z  \\
\end{matrix} \right)
\end{align}
Since $Z\bar{Z}$ and $\bar{Z}Z$ share the same eigenvalues ($\{P,Q\}=0$ thus $P$ flips the sign of eigenvalues of $Q$, but after squaring these two eigenvalues with opposite signatures become the same), and there is no internal structure in $Z$ (in this case $R$ is a symmetry of $Q$, $[R,Q]=0$, but $R^2=1$, thus $R$ cannot provide any further degeneracy), we obtain that $H$ has a two-fold degeneracy.
\\
\\
Similar to the previous $N \bmod 8=2$ case, we can take the operator $Q$ and $H$ as the whole matrices instead of blocks. For $H$, we notice that the transposing operations make the exchange of these two sectors. However, the symmetric matrix statement is basis-dependent. Formally, similar to the quaternion Hermitian case, we can extend the definition of the symmetric matrix by the following. Define
\begin{align}
\Omega'=\left( \begin{matrix}
   0 & 1  \\
   1 & 0  \\
\end{matrix} \right)
\end{align}
and we could see that, a matrix $M$ is symmetric real (or symmetric Hermitian) if and only if $M^\dagger=M$ and $M^T\Omega'=\Omega' M$ (where $\Omega'$ means the basis changing over two sectors). We can easily check that $Q$ satisfies this condition. Thus $Q^2=H$ must satisfy. Thus we conclude that the total matrix $H$ in a subset of GOE symmetry class (with positivity constraint).
\end{itemize}
Although from the symmetric point of view, the Hamiltonian of $\mathcal{N}=1$ model should be classified in the subsets of standard Dyson ensembles. But what the subset exactly is? In fact, the special structure of the squaring from $Q$ to $H$ will change the distribution of the eigenvalues from Gaussian to Wishart-Laguerre \cite{Wishart,statistical,MIT} (Although there are some differences in the powers of terms in the eigenvalue distributions.) We will roughly call them as LOE/LUE/LSE, as has been used in the random matrix theory research. Some more details will be summarized in the appendix \ref{Dist}.
\\
\\
However, the difference in the details of the distribution, beyond numerical tests of the distribution function of the one point-eigenvalues, will not be important in some physical tests, such as spectral form factors and level statistics (e.g., the Wigner surmise). The reason could be given as follows. From the supercharge point of view, because $Q$ is in the Altland-Zirnbauer distribution with non-trivial $\tilde{\alpha}$ (see appendix \ref{Dist}), the squaring operation will not change the level statistics such as Wigner surmise and spectral form factors (which could also be verified by numerics later). From the physical point, as is explained in \cite{You:2016ldz}, the details of distribution (even if not Gaussian), cannot change the universal properties of symmetries.
\\
\\
Finally, we can summarize these statements in the following classification table (the degeneracies have been already calculated in \cite{Fu:2016vas}),
\begin{center}
\begin{tabular}{ c | c | c | c | c | c   }
$N \bmod 8$ & Deg. & RMT & Block & Type & Level stat.\\
\hline
0 & 2 & $\text{LOE}$ & $\left( \begin{matrix}
   AA^T & 0  \\
   0 & A^TA  \\
\end{matrix} \right)\text{ }A \text{ real}$
&$\mathbb{R}$& GOE\\
2 & 4 & $\text{LSE}$ &$\left( \begin{matrix}
   -Y\bar{Y} & 0  \\
   0 & -\bar{Y}Y  \\
\end{matrix} \right)\text{  }Y \text{ complex skew-symmetric}$
&$\mathbb{H}$& GSE\\
4 & 4 & $\text{LSE}$ &$\left( \begin{matrix}
   BB^\dagger & 0  \\
   0 & B^\dagger B  \\
\end{matrix} \right)\text{  }B \text{ Hermitian quaternion}$ &$\mathbb{H}$& GSE\\
6 & 2 & $\text{LOE}$ &$\left( \begin{matrix}
   Z\bar{Z} & 0  \\
   0 & \bar{Z}Z  \\
\end{matrix} \right)\text{  }Z \text{ complex symmetric}$ &$\mathbb{R}$& GOE\\
\end{tabular}
\end{center}
For our further practical computational usage, we may summarize the following table for different $N$s in the supersymmetric SYK random matrix correspondence. As we show in the next section, for $N \ge 14$, these theoretical considerations perfectly fit the level statistics.
\begin{center}
\begin{tabular}{ c | c | c | c | c | c | c| c| c| c | c }
$N$ & 10 & 12 & 14 & 16 &18 & 20 & 22 & 24 & 26 & 28 \\
\hline
RMT & LSE & LSE & LOE & LOE& LSE& LSE & LOE & LOE & LSE & LSE\\
Universal Stat.& GSE & GSE & GOE & GOE & GSE & GSE & GOE & GOE & GSE & GSE
\end{tabular}
\end{center}
\footnotetext{We say a matrix $M$ is a quaternion Hermitian matrix if and only if \[M=\left( \begin{matrix}
   A+iB & C+iD  \\
   -C+iD & A-iB  \\
\end{matrix} \right)\] for some real $A,B,C,D$ in a basis, and $A$ is symmetric while $B,C,D$s are skew-symmetric. There is an equivalent definition that, defining
\[\Omega=\left( \begin{matrix}
   0 & 1  \\
   -1 & 0  \\
\end{matrix} \right)\]
then $M$ is a quaternion Hermitian matrix if and only if $M^\dagger=M$ and $M\Omega=\Omega M$. Thus it is shown directly that if $M$ is quaternion Hermitian then $(M M^\dagger)^\dagger=M M^\dagger$ and $M M^\dagger \Omega= M (M \Omega)=M \Omega M=\Omega M^2=\Omega M M^\dagger$, thus $M M^\dagger=M^2=M^\dagger M$ is still a quaternion Hermitian matrix.
}

\section{Exact Diagonalization}\label{data}
In this part, we will present the main results from numerics to test the random matrix theory classification in the previous investigations. One can diagonalize the Hamiltonian exactly with the representation of the Clifford algebra by the following. For operators acting on $N_d=N/2$ qubits, one can define
\begin{align}
  & {{\gamma }_{2\zeta -1}}=\frac{1}{\sqrt{2}}\left( \prod_{p=1}^{{{N}_{d}}-1}\sigma _{p}^{z} \right)\sigma _{{{N}_{d}}}^{x} \nonumber\\
 & {{\gamma }_{2\zeta }}=\frac{1}{\sqrt{2}}\left( \prod _{p=1}^{{{N}_{d}}-1}\sigma _{p}^{z} \right)\sigma _{{{N}_{d}}}^{y}
\end{align}
where $\sigma_p$ means standard Pauli matrices acting on the $p$-th qubit, tensor-producting the identity matrix on the other parts, and $\zeta=1,2,...... ,N_d$. This construction is a representation of the Clifford algebra
\begin{align}
\left\{ {{\gamma }_{a}},{{\gamma }_{b}} \right\}={{\delta }_{ab}}
\end{align}
And one can exactly diagonalize the Hamiltonian by replacing the Majonara fermions with gamma matrices to find the energy eigenvalues. Thus, all quantities are computable by brute force in the energy eigenstate basis.
\\
\\
The main results of the following investigation would be the following. In the density of the supercharge eigenstates and energy eigenstates in the supersymmetric SYK model, the behavior is quite different, but coincides with our estimations from the random matrix theory classification: the spectral density of the supercharge $Q$ shows clearly the information about extended ensembles from Altland-Zirnbauer theory, and the spectral density of energy $H$ shows a clear Marchenko-Pastur distribution from the statistics of Wishart-Laguerre. Moreover, because both $Q$ and $H$ both belong to the universal level statistical class for GOE, GUE, and GSE, the numerics from Wigner surmise and spectral form factor will show these eight-fold features directly.
\subsection{Density of states}
\begin{figure}[t]
  \centering
  \includegraphics[width=0.3\textwidth]{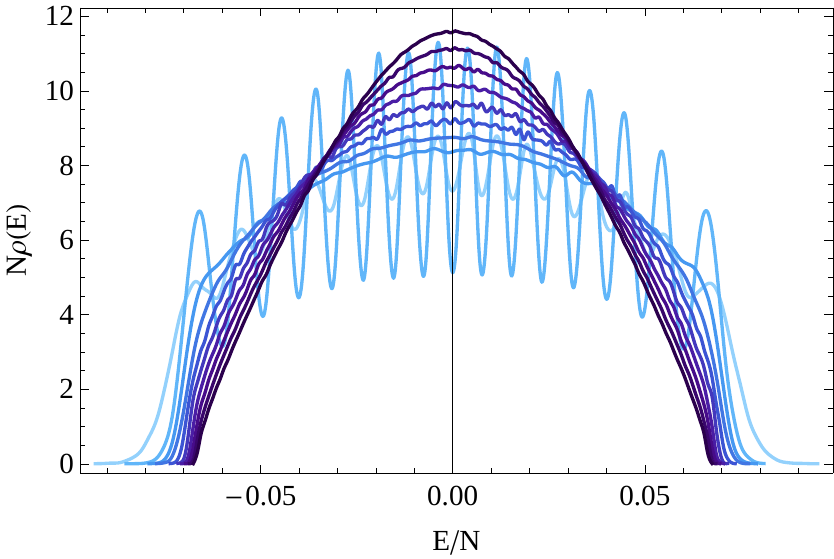}~~~
  \includegraphics[width=0.3\textwidth]{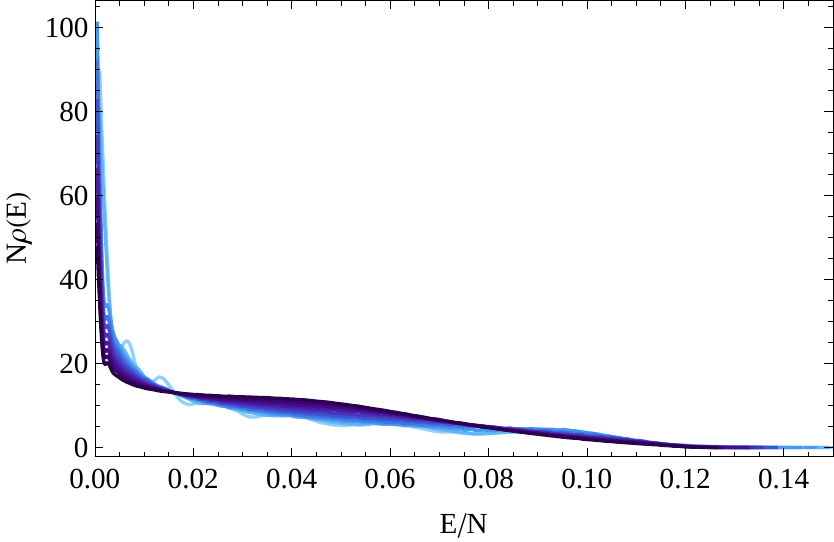}~~~
  \includegraphics[width=0.3\textwidth]{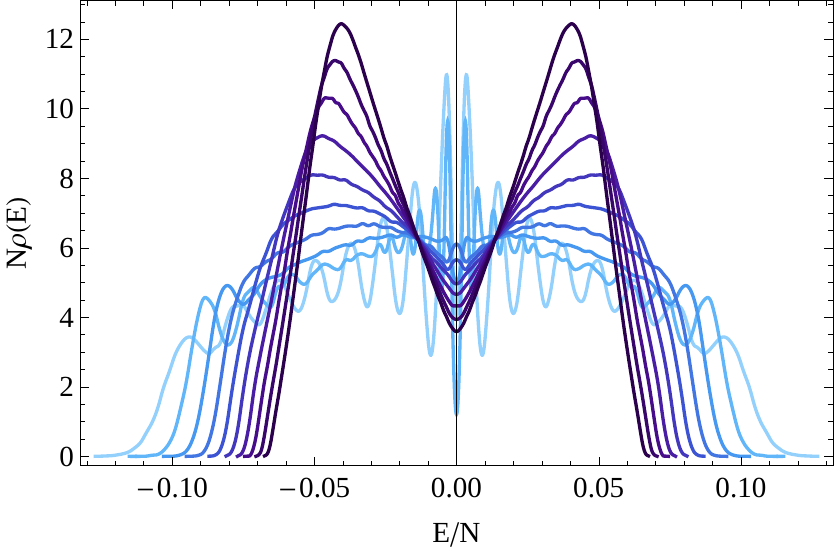}~~~
  \caption{\label{rho} The density of states for the original SYK model Hamiltonian (left), the supersymmetric SYK Hamiltonian (middle), and the supersymmetric SYK supercharge operators treated as Hamiltonian (right) by exact diagonalization. The densities of states from $N=10$ to $N=28$ are plotted in colors from light blue to dark blue. The eigenvalues have been rescaled by $E(Q)/NJ$, while the density of states has also been rescaled to match the normalization that the integration should be 1.}
\end{figure}
The plots for the density of states in the SYK model and its supersymmetric extension are shown in Figure \ref{rho} for comparison. For each realization of random Hamiltonian, we compute all eigenvalues. After collecting a large number of samples, one can plot the histograms for all samples as the function $\rho(E)$. For the density of states in the SYK model, in small $N$, tiny vibrations are contained, while in the large $N$ the distribution will converge to a Gaussian distribution besides the
small tails. However, in the supersymmetric SYK model, the energy eigenvalue structure is totally different. All energy eigenvalues are larger than zero because $H=Q^2> 0$. Because of supersymmetry, the lowest energy eigenvalues will approach zero for large $N$, and the figure will come to a convergent distribution. The shape of this distribution matches the eigenvalue distribution of Wishart-Laguerre, which is the Marchenko-Pastur distribution \cite{lec} in the large $N$ limit. For the supercharge matrices, as $N$ becomes larger the curve acquires a dip at zero, which is a clear feature for extended ensembles and could match the averaged
density of eigenvalues of random matrices in CI, DIII \cite{AZ1997}
and chiral \cite{Jackson:1996xt} ensembles at large $N$.
\\
\\
For numerical details, we compute $N=10$ (40000 samples), $N=12$ (25600 samples), $N=14$ (12800 samples), $N=16$ (6400 samples), $N=18$ (3200 samples), $N=20$ (1600 samples), $N=22$ (800 samples), $N=24$ (400 samples), $N=26$ (200 samples), and $N=28$ (100 samples). The results for original SYK model perfectly match the density of states obtained in previous works (e.g., \cite{Maldacena:2016hyu,Cotler:2016fpe}).
\subsection{Wigner surmise}\label{sec:wigner}
\begin{figure}[t]
  \centering
  \includegraphics[width=0.6\textwidth]{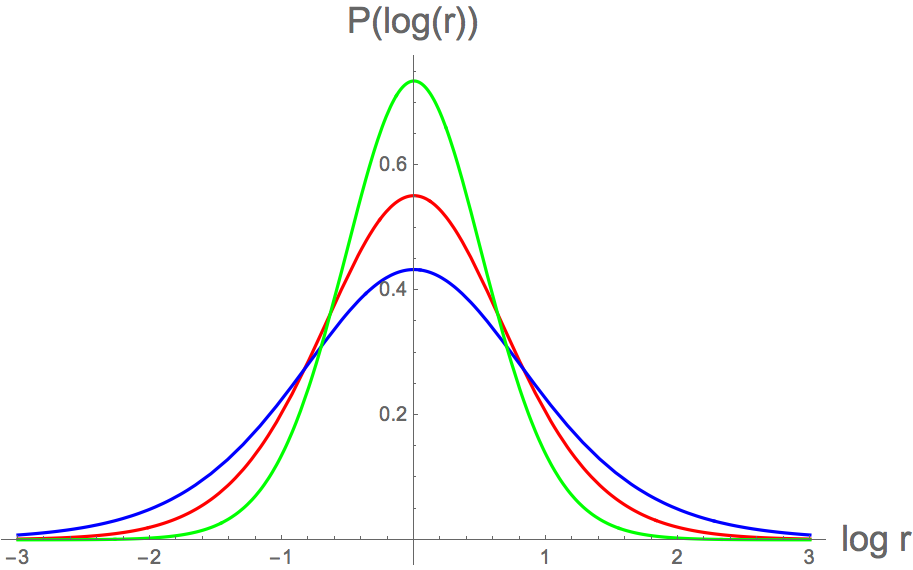}
  \caption{\label{wignerstandard} The theoretical Wigner surmises for three different standard ensembles. The lower (blue), middle (red) and higher (green) curves are corresponding to GOE, GUE and GSE universal class respectively.}
\end{figure}
There exists a practical way to test if random matrices from a theory are from some specific ensembles. For a random realization of the Hamiltonian, we have a collection of energy eigenvalues $E_n$. If we arrange them in ascending order $E_n<E_{n+1}$, we define, $\Delta E_n= E_n- E_{n-1}$ to be the level spacing, and we compute the ratio for the nearest neighbourhood spacing as $r_n=\Delta E_n/\Delta E_{n+1}$. For matrices from the standard Dyson ensemble, the distribution of level spacing ratio satisfies the Wigner-Dyson statistics\cite{wignerSurmise}) (which is called the \emph{Wigner surmise}
\begin{align}
p(r)=\frac{1}{Z}\frac{{{(r+{{r}^{2}})}^{\tilde{\beta} }}}{{{(1+r+{{r}^{2}})}^{1+3\tilde{\beta} /2}}}
\end{align}
for GOE universal class, $\tilde{\beta}=1$, $Z=8/27$; for GUE universal class, $\tilde{\beta}=2$, $Z=4\pi/(81\sqrt{3})$; for GSE universal class, $\tilde{\beta}=4$, $Z=4\pi/(729\sqrt{3})$ (In fact, these are labels for the field of representation. See appendices for more details). Practically we often change $r$ to $\log r$, and the new distribution after the transformation is $P(\log r)=r p(r)$. Standard Wigner surmises are shown in the Figure \ref{wignerstandard}. \cite{You:2016ldz} has computed the nearest-neighbor level spacing distribution of the SYK model, which perfectly matches the prediction from the eight-fold classification.
\\
\begin{figure}[t]
  \centering
  \includegraphics[width=\textwidth]{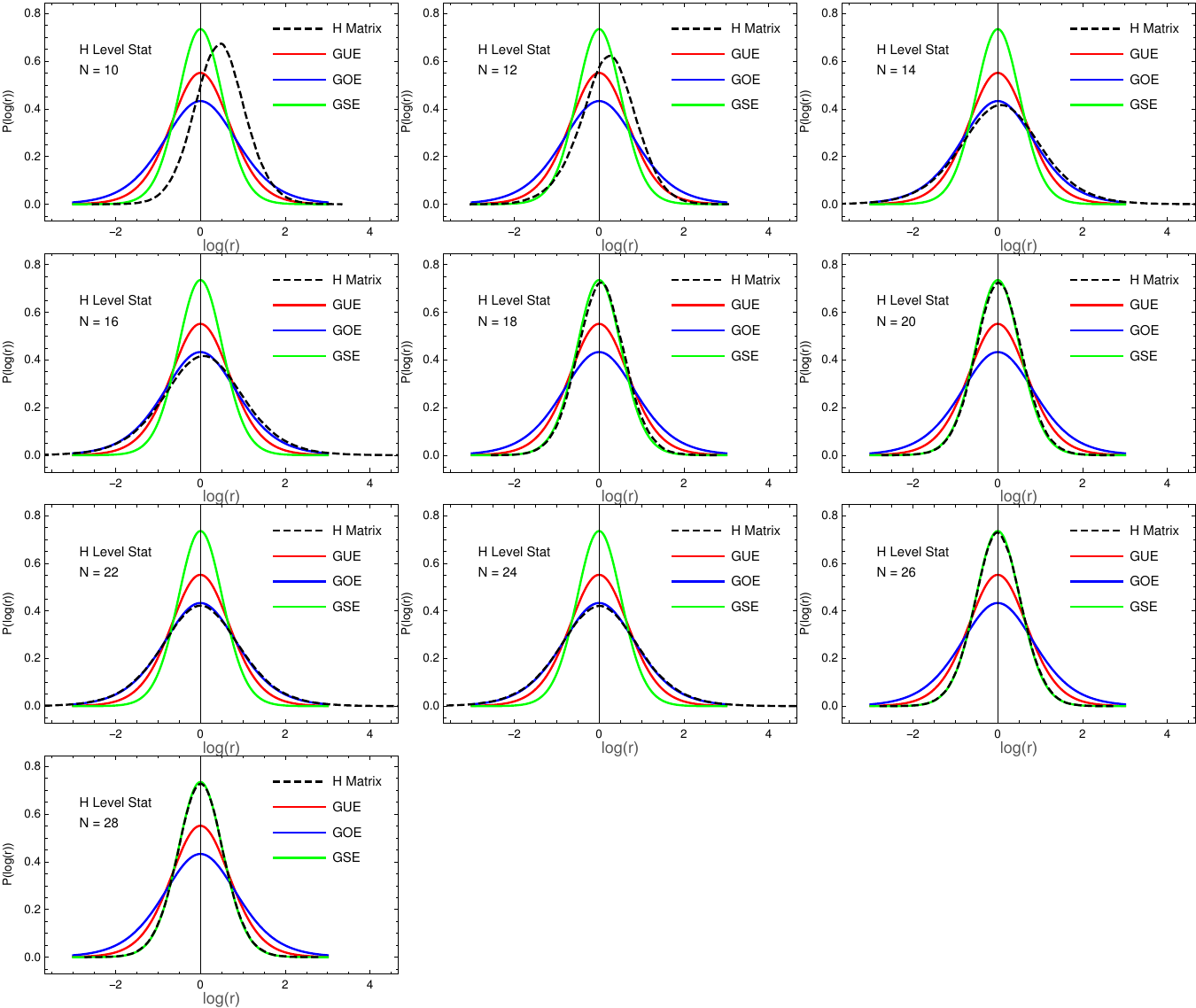}\vspace{-3.8cm}
  \flushright
  \begin{minipage}{0.62\textwidth}
    \caption{\label{wignersusy} The nearest-neighbor level spacing distribution for the Hamiltonian of the $\mathcal{N}=1$ supersymmetric SYK model for different $N$. The lower (blue), middle (red), and higher (green) curves are theoretical predictions of Wigner surmises from GOE, GUE, and GSE, respectively. The black dashed curves are distributions for all $r$s from a large number of samples. 
    }
  \end{minipage}
\end{figure}
\\
What is the story for the $\mathcal{N}=1$ supersymmetric SYK model? A
numerical investigation shows a different correspondence for the eight-fold classification, which is given by Figure \ref{wignersusy}. One can clearly see the new correspondence in the eight-fold classification for supersymmetric SYK models, as has been predicted in the previous discussions.
\begin{figure}[t]
  \centering
  \includegraphics[width=\textwidth]{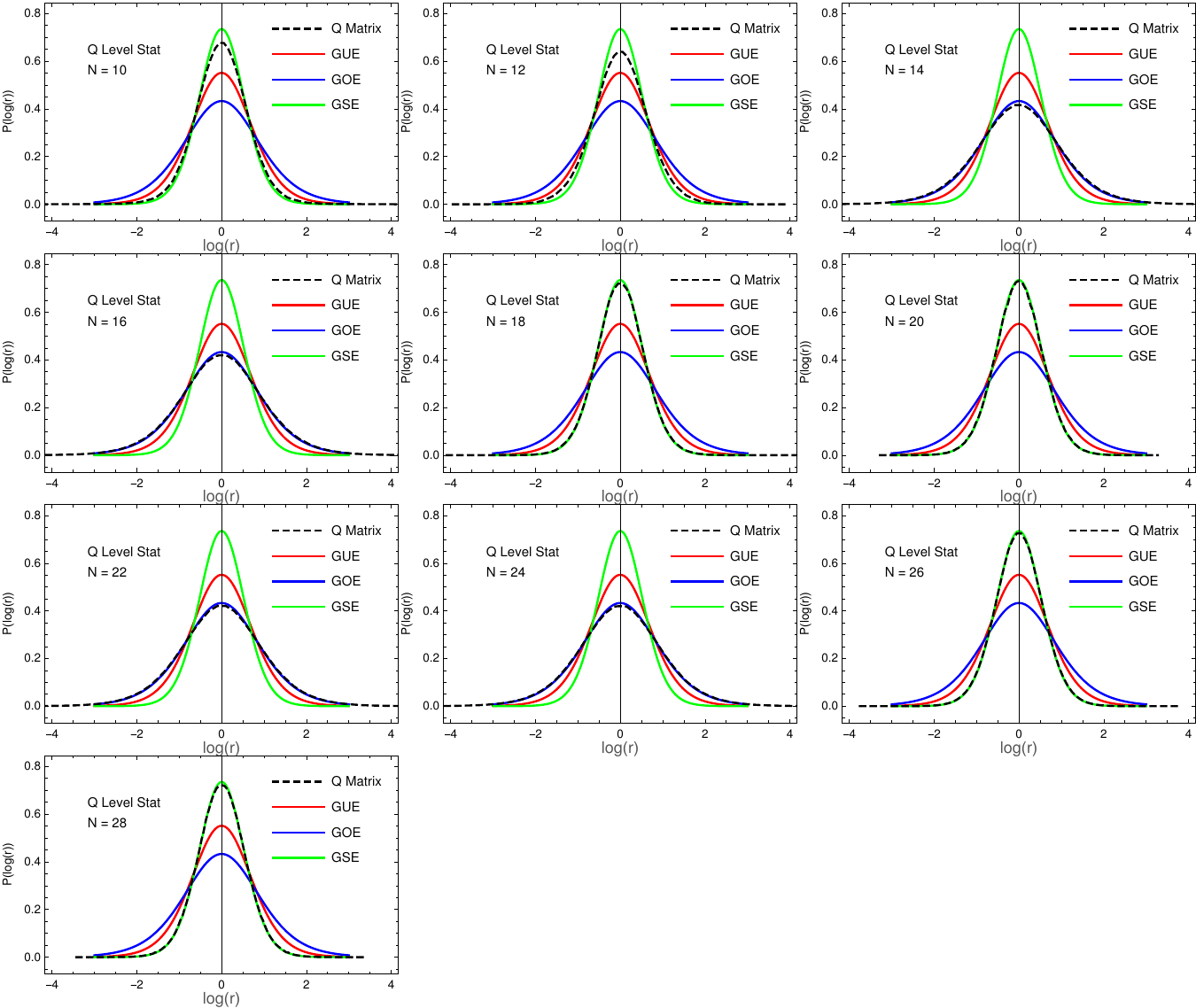}\vspace{-3.8cm}
  \flushright
  \begin{minipage}{0.62\textwidth}
    \caption{\label{fig:qstat} The nearest-neighbor level spacing distribution for the supercharge matrix $Q$ of $\mathcal{N}=1$ supersymmetric SYK model for different $N$.  The lower (blue), middle (red), and higher (green) curves are the theoretical prediction of Wigner surmises from GOE, GUE, and GSE, respectively. The black dashed curves are distributions for all $r$s from a large number of samples.  }
  \end{minipage}
\end{figure}
\\
\\
Some comments should be given in this prediction. Firstly, one has some subtleties in obtaining correct $r$s. Considering there are two different parities in the SYK Hamiltonian ($F \text{ mod } 2$), each group of parity should only appear once in the statistics of $r_n$. For $N \text{ mod } 8= 0,4$ in SYK, the particle-hole operator $P$ maps each sector to itself, thus if we take all $r_n$ the distribution will be ruined, serving as a many-body-localized distribution (the Poisson distribution). For $N \text{ mod } 8= 2,6$ in the SYK model, the particle-hole operator $P$ maps even and odd parities to each other, and one can take all possible $r$s in the distribution because all fermionic parity sectors are degenerate. Similar things are observed for all even $N$ in the supersymmetric SYK model. As we mentioned before, the reason is that the supercharge $Q$ is a symmetry of $H$, which always changes the particle number because it is an odd-point coupling term. Moreover, the standard ensemble behavior is only observed for $N \ge 14$, and for small enough $N$s, we have no clear correspondence. Similar things happen for original SYK model, where the correspondence works only for $N \ge 5$, because there is no
thermalization if $N$ is too small \cite{You:2016ldz}. However, the threshold for obtaining a standard random matrix from the $\mathcal{N}=1$ supersymmetric extension is much larger.
\\
\\
In Section~\ref{sec:RMTQ}, we argued that the supercharge operator $Q$ in $\mathcal{N}=1$ supersymmetric SYK theory are also random matrices in some extended ensembles \cite{AZ1997,Zirnbauer1996}. We compute the level statistics of $Q$ and compare it with the Wigner surmises of three standard Dyson ensembles in cases with different $N$. The result is presented in Figure~\ref{fig:qstat}. We see the level statistics of $Q$ matrices match the same ensembles as the corresponding Hamiltonian. This result confirms the relationship between $Q$'s random matrix ensemble and that of the corresponding $H$. That we do not see extended ensemble in the $Q$'s level statistics because the level statistic does not see all the information in the ensembles.

\subsection{Spectral form factors}
\begin{figure}[t]
  \centering
  \includegraphics[width=0.9\textwidth]{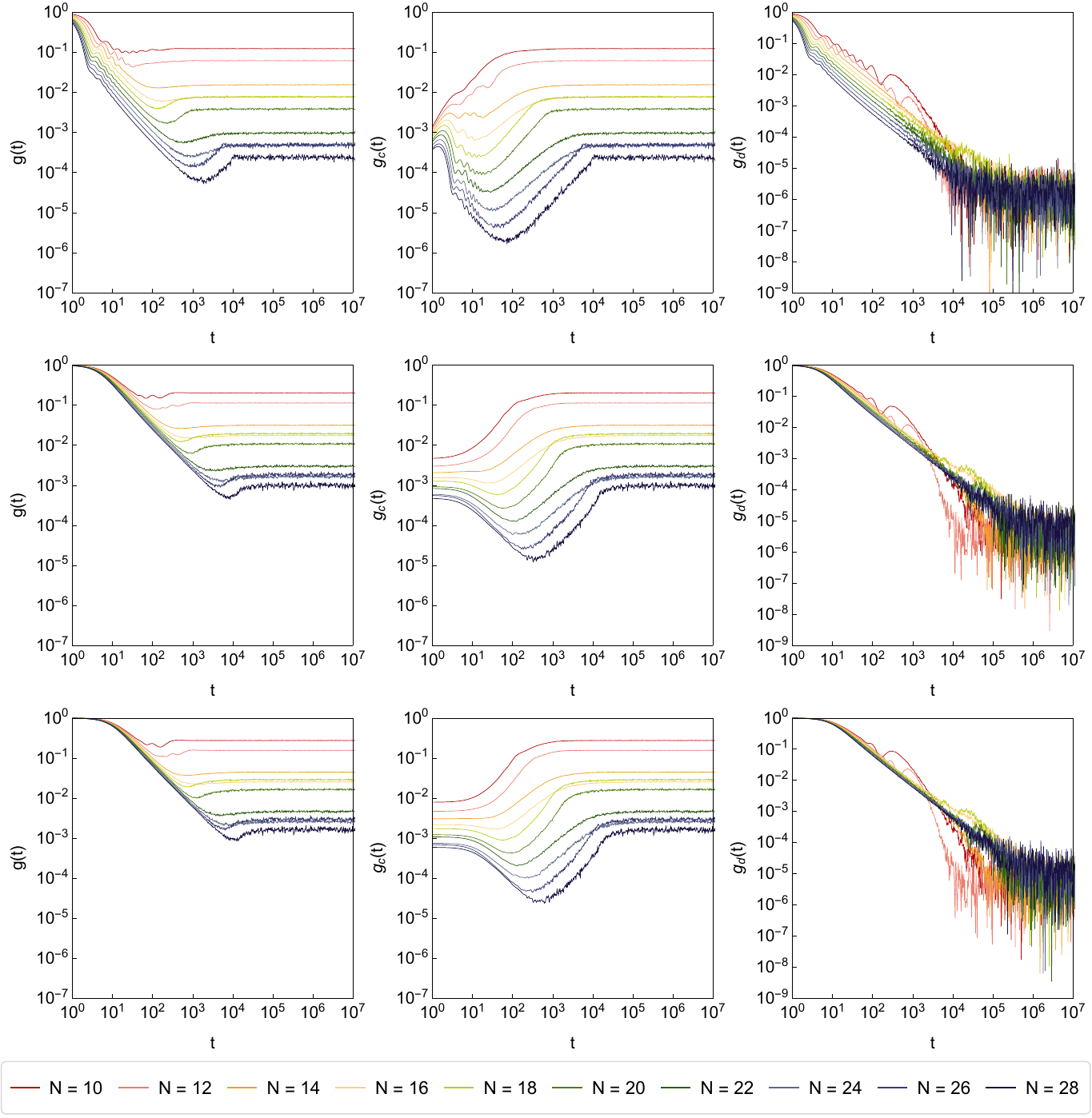}
  \caption{\label{spec_SUSY}  The spectral form factors $g(t)$, $g_c(t)$ and $g_d(t)$ in the supersymmetric SYK model with $J_{\mathcal{N}=1}=1$, $\beta=0,5,10$ respectively.}
\end{figure}
\begin{figure}[t]
  \centering
  \includegraphics[width=0.9\textwidth]{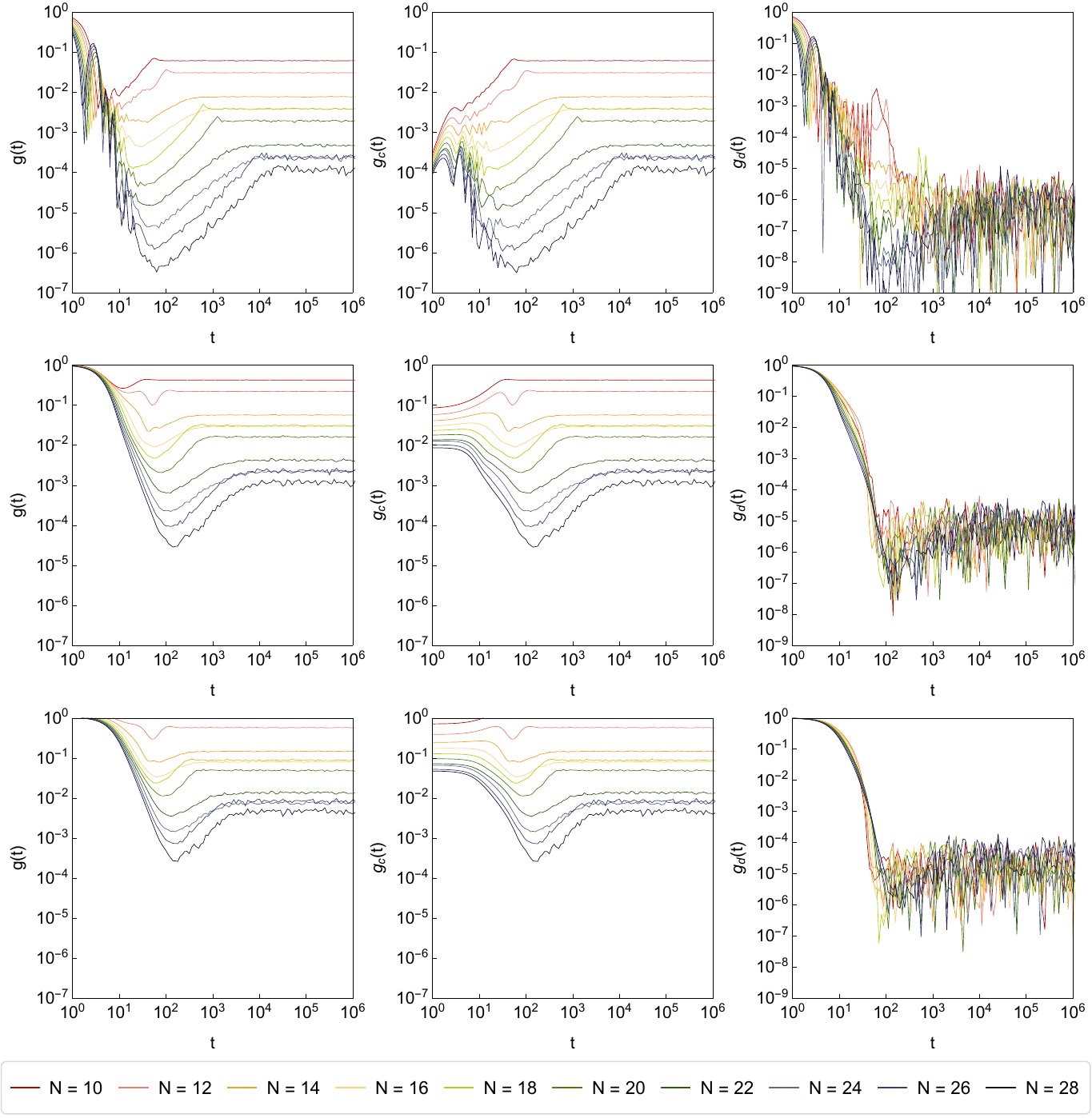}
  \caption{\label{fig:QSFF}  The ``spectral form factors'' $g(t)$, $g_c(t)$ and $g_d(t)$ in the supersymmetric SYK model, treating the supercharge matrix as the Hamiltonian, with $J_{\mathcal{N}=1}=1$, $\beta=0,5,10$ respectively.}
\end{figure}
Before presenting the numeric results of spectral form factors, we will review the discreteness of spectra and the spectral form factor following \cite{Cotler:2016fpe}. For a quantum mechanical system, the partition function
\begin{align}
Z(\beta )=\text{Tr}({{e}^{-\beta H}})
\end{align}
could be continued as
\begin{align}
Z(\beta ,t)=Z(\beta +it)=\text{Tr}({{e}^{-\beta H-iHt}})
\end{align}
The analytically continued partition function $Z(\beta,t)$ is an important quantity to understand a discrete energy spectrum. Typically, people will compute the time average to understand the late time behavior, but for $Z(\beta,t)$, it vibrates near zero at the late time, and the time average should be zero. Thus, we often compute ${{\left| \frac{Z(\beta ,t)}{Z(\beta )} \right|}^{2}}$. For a discrete energy eigenvalue spectrum, we have
\begin{align}
{{\left| \frac{Z(\beta ,t)}{Z(\beta )} \right|}^{2}}=\frac{1}{Z{{(\beta )}^{2}}}\sum\limits_{m,n}{{{e}^{-\beta ({{E}_{m}}+{{E}_{n}})}}{{e}^{i({{E}_{m}}-{{E}_{n}})t}}}
\end{align}
It's hard to say anything general directly for a general spectrum, but one can use the long-term average
\begin{align}
\frac{1}{T}\int_{0}^{T}{{{\left| \frac{Z(\beta ,t)}{Z(\beta )} \right|}^{2}}dt}=\frac{1}{Z{{(\beta )}^{2}}}\sum\limits_{E}{n_{E}^{2}{{e}^{-2{{\beta }_{E}}}}}
\end{align}
for large enough $T$ ($n_E$ means the degeneracy). For a non-degenerate spectrum, it should have a simple formula
\begin{align}
{{\left| \frac{Z(\beta ,t)}{Z(\beta )} \right|}^{2}}=\frac{Z(2\beta )}{Z{{(\beta )}^{2}}}
\end{align}
However, for a continuous spectrum, the quantity has vanishing long-term average. Thus, the quantity should be an important criterion to detect the discreteness. In this paper, we will use a similar quantity, which is called the spectral form factor
\begin{align}
  & g(t,\beta )=\frac{\left\langle Z(\beta +it)Z(\beta -it) \right\rangle }{{{\left\langle Z(\beta ) \right\rangle }^{2}}} \nonumber\\
 & {{g}_{d}}(t,\beta )=\frac{\left\langle Z(\beta +it) \right\rangle \left\langle Z(\beta -it) \right\rangle }{{{\left\langle Z(\beta ) \right\rangle }^{2}}} \nonumber\\
 & {{g}_{c}}(t,\beta )=g(t,\beta )-{{g}_{d}}(t,\beta )=\frac{\left\langle Z(\beta +it)Z(\beta -it) \right\rangle -\left\langle Z(\beta +it) \right\rangle \left\langle Z(\beta -it) \right\rangle }{{{\left\langle Z(\beta ) \right\rangle }^{2}}}
\end{align}
In the SYK model, these quantities will have similar predictions with the Hamiltonian replaced by random matrix from some specific given Dyson ensembles. For example, for a given realization $M$ from a random matrix ensemble with large $L$, we have the analytically continued partition function
\begin{align}
{{Z}_{\text{rmt}}}(\beta ,t)=\frac{1}{{{\mathcal{Z}}_{\text{rmt}}}}\int{d{{M}_{ij}}}\exp \left( -\frac{L}{2}\operatorname{Tr}({{M}^{2}}) \right)\text{Tr(}{{e}^{-\beta M-iMt}}\text{)}
\end{align}
where
\begin{align}
{{\mathcal{Z}}_{\text{rmt}}}=\int{d{{M}_{ij}}}\exp \left( -\frac{L}{2}\operatorname{Tr}({{M}^{2}}) \right)
\end{align}
The properties of spectral form factors given by random matrix theory, $g_\text{rmt}(t)$, have been studied in \cite{Cotler:2016fpe}. There are three specific periods in $g_\text{rmt}(t)$. In the first period, the spectral form factor will quickly decay to a minimal until \emph{dip time} $t_d$. Then after a short increasing (the \emph{ramp}) towards a \emph{plataeu time} $t_p$, $g_\text{rmt}(t)$ will arrive at a constant plataeu. This pattern is extremely similar with SYK model. Theoretically, in the early time (before $t_d$), $g(t)$ should not obtained by $g_\text{rmt}(t)$ because of different initial dependence on energy, while in the late time these two systems are conjectured to be coincide \cite{Cotler:2016fpe}.
\\
\\
With the data of energy eigenvalues, one could compute the spectral form factors, which have been shown in Figure \ref{spec_SUSY} for the supersymmetric SYK model. We perform the calculation for three different functions $g(t)$, $g_d(t)$ and $g_c(t)$ with $\beta=0, 5, 10$ and several $N$s. Clear patterns similar to random matrix theory predictions are shown in these numerical simulations. One could directly see the dip, ramp, and plateau periods. For small $\beta$s, there exist some small vibrations in the early time, while for large $\beta$, this effect disappears. The function $g_d$ is strongly vibrating because we have only a finite number of samples. One could believe that the infinite number of samples will cancel the noisy randomness of the curves.
\\
\\
A clear eight-fold correspondence has been shown in the spectral form factor. Near the plateau time of $g(t)$, one should expect roughly a smooth corner for GOE-type, a kink for GUE-type, and a sharp peak for GSE-type. Thus, we observe roughly the smooth corners for $N=14,16,22,24$, while the sharp peaks for $N=18,20,26,28$ (although the peaks look not very clear because of finite sample size). For $N=10,12$, as shown in Figure \ref{wignersusy}, there is no clear random matrix correspondence because $N$ is too small. Thus we only observe some vibrations near the plateau time.
\\
\\
We also perform a similar test on the supercharge $Q$, plotted in Figure \ref{fig:QSFF}. In Section~\ref{sec:wigner}, we numerically tested the nearest neighbor level statistics of $Q$, which matches perfectly the statistics of the corresponding $H$. The spectral form factors of $Q$ are slightly different from those of $H$, yet they show exactly the same eight-fold behavior.

\subsection{Dip time, plateau time and plateau height}
\begin{figure}[t]
  \centering
  \includegraphics[width=1.0\textwidth]{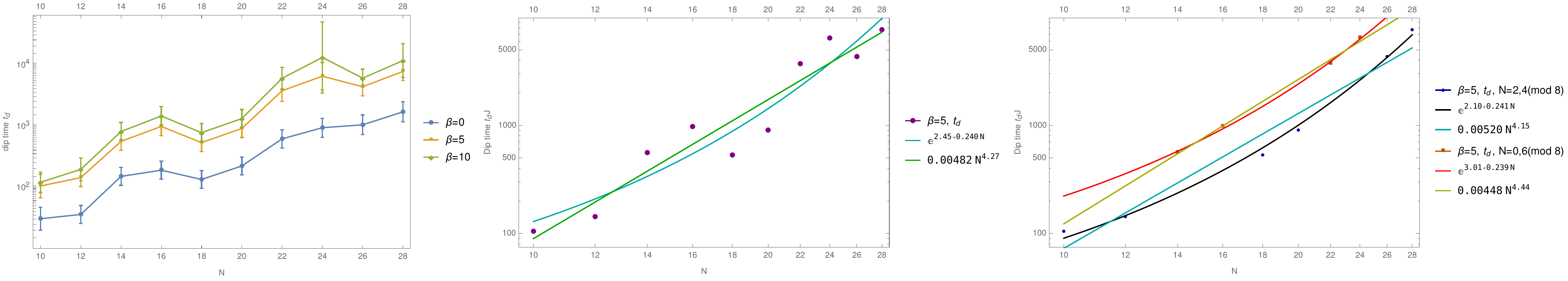}
  \caption{\label{diptime} The dip time $t_d$ for the supersymmetric SYK model. In the left figure, we evaluate three different temperatures and compute the dip time with respect to $N$, where the error bar is given as the standard deviation when evaluating $t_d$ because of large noise is around the minimal point of $g(t)$. In the middle figure, we fit the dip time by polynomials and exponential functions for $t_d(N)$ at the temperature $\beta=5$. In the right figure, we separately fit the dip time for two different random matrix classes with the same temperature $\beta=5$ and the same fitting functions.}
\end{figure}

\begin{figure}[h]
  \centering
  \includegraphics[width=1.0\textwidth]{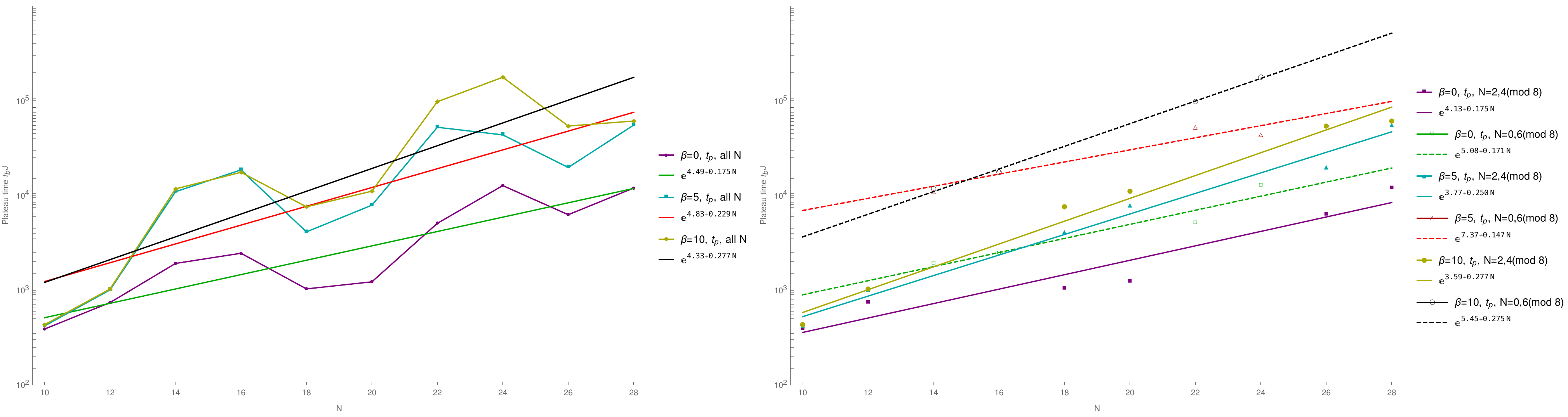}
  \caption{\label{platime} The plateau time $t_p$ for the supersymmetric SYK model. We choose three different temperatures and evaluate the plateau time with respect to $N$, and we use the exponential function to fit $t_p(N)$. In the left figure we use all $N$s, while in the right figure we separately fit two different random matrix classes.}
\end{figure}

\begin{figure}[h]
  \centering
  \includegraphics[width=1.0\textwidth]{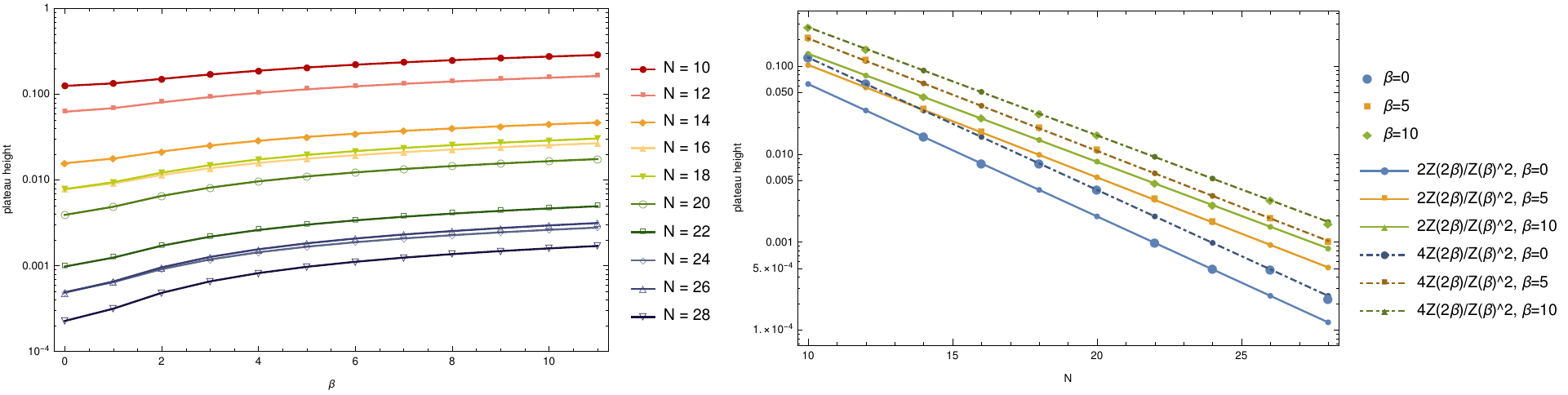}
  \caption{\label{platheight} The plateau height $g_p$ for the supersymmetric SYK model. In the left figure, we choose several temperatures and fix $N$ in each curve, while in the right, we fix $\beta$ and evaluate $g_p(N)$.}
\end{figure}

More quantitative data could be read off from the spectral form factors. In Figure \ref{diptime}, Figure \ref{platime} and Figure \ref{platheight} we present our numerical results for dip time $t_d$ of $g(t)$, plateau time $t_p$ of $g(t)$, and plateau height $g_d$ of $g_c(t)$ respectively. For numerical technics, we choose the linear fitting in the ramp period, and the plateau is fitted by a straight line parallel to the time axis. The dip time is read off as the averaged minimal point at the end of the dip period, and the error bar could be computed as the standard deviation.
\\
\\
It is claimed in \cite{Cotler:2016fpe} that polynomial and exponential fitting could be used to interpret the dip time as a function of $N$ with fixed temperature. We apply the same method to the supersymmetric extension. However, we find that in the supersymmetric extension, the fitting is much better if we fit the GOE-type group ($N \bmod 8=0,6$) and the GSE-type group ($N \bmod 8=2,4$) separately. On the other hand, although we cannot rule out the polynomial fitting, the fitting effect of the exponential function is relatively better. On the exponential fittings with respect to different degeneracy groups, the coefficients before $N$ are roughly the same ($0.24N$ for $\beta=5$) while the overall constants are different. That indicates that the eight-fold degeneracy class or random matrix class might influence the overall factors in the dip time exponential expressions.
\\
\\
One could also read off the plateau time and exponentially fit the data. Similar to dip time, we could also separately fit the plateau time with respect to two different random matrix classes, and one could find a difference in the overall factors of these two groups, while the coefficients before $N$ are the same. There is a non-trivial check here. Theoretically from random matrix theory one can predict that the plateau time scales like $t_p\sim e^{S(2\beta)}$ \cite{Cotler:2016fpe}. In the large $\beta$ limit, the entropy should be roughly the ground state entropy. Analytically, the entropy is predicted by $S(\beta=\infty)=N s_0=0.275 N$. Now check the largest $\beta$ we take ($\beta=10$). We can read off the entropy by $0.277N$ (GSE-type), $0.275N$ (GOE-type), or $0.277N$ (two groups together), which perfectly matches our expectations.
\\
\\
For the plateau height, one can clearly see an eight-fold structure. From the previous discussion, we obtain that the plateau height should equal to $Z(2\beta)/Z(\beta)^2$ times a contribution from the degeneracy, which is clearly shown in the figure. For $N =14,16,22,24$ (GOE-type), the degeneracy is two. Thus points should be on the lower line, while for $N=18,20,26,28$ (GSE-type), the degeneracy is four. Thus points should be on the upper line. These observations match the prediction from random matrix theories.

\section{Conclusion and outlook}\label{conclu}
In this paper, we use analytic arguments and numerical evidence to explore the supersymmetric constraints on the random matrix theory symmetry class. We focus on the $\mathcal{N}=1$ supersymmetric SYK model, a supersymmetric generalization of nonlocal-coupled Majonara fermions with similar chaotic behavior for a two-dimensional quantum black hole.
\\
\\
Use the direct classification from random matrix theory, we show that for $\mathcal{N}=1$ supersymmetric SYK model has different behavior for $N \text{ mod } 8$ structure. These arguments might be made to be more general: supersymmetry could directly change the universal class of Hamiltonian (GOE/GUE/GSE) by classifying the symmetry class of the supercharge, where combinations of Witten index and anti-unitary operators will make some new anti-unitaries; On the other hand, the quadratic structure of the Hamiltonian will change the original type of distribution from Gaussian to Wishart-Laguerre. These points may happen for generic supersymmetric statistical physics models.
\\
\\
We also use the numerical method, exact diagonalization, to confirm the random matrix theory classification on the Hamiltonian and the supercharge of the supersymmetric SYK model. It is clear that if we check the spectrum density, the supercharge $Q$ shows a clear feature from the one-point function of extended random matrix theory ensembles, while the Hamiltonian shows a feature of the quadratic semi-circle (Marchenko-Pastur). However, for level statistics (e.g., Wigner surmise and spectral form factor), the universal class GSE/GOE could capture important physical features, and the new eight-fold rule could be verified.
\\
\\
Several future directions could be investigated. Firstly, one may consider higher supersymmetry constraints on the SYK model, such as $\mathcal{N}=2$ generalization. Many thermodynamical and field theory properties of higher supersymmetric SYK theory are non-trivial, and it might be interesting to connect these properties to random matrix theory. Moreover, to understand the spectral form factor with supersymmetric constraints, one could also try to study superconformal field theory partition functions at late time. Finally, introducing supersymmetries in the symmetry classification of phases in the condensed matter theory will bring more understanding at the boundary of condensed matter and high energy physics. We leave these interesting possibilities for future works.


\section*{Acknowledgments}
We thank Xie Chen, Kevin Costello, Liam Fitzpatrick, Davide Gaiotto, Yingfei Gu, Nicholas Hunter-Jones, Alexei Kitaev, Andreas Ludwig, Evgeny Mozgunov, Alexandre Streicher for valuable discussions. We thank Takuya Kanazawa for comments on the draft. JL is deeply grateful to Guy Gur-Ari for communications on the symmetry of the original and supersymmetric SYK models. TL, JL, YX, and YZ are supported by graduate student programs of the University of Nebraska, Caltech, Boston University, and Perimeter Institute.
\appendix
\section{Review on the Altland-Zirnbauer theory}\label{AZ}
In this appendix, we make a brief review of the Altland-Zirnbauer theory (e.g., see \cite{Zirnbauer1996,AZ1997}) that brings Hamiltonians to ten different random matrix classes. In a physical system, symmetries can appear, which is given by a group $G$, then the space of physical states is a projective representation of the symmetry group. A fundamental question we can ask is, what is the most general type of Hamiltonian the system can have.
\\
\\
We may visit the simplest example to get some intuitions. The action of an element of $G$ on the Hilbert space $V$ can be either unitary or antiunitary. Thus there is a homomorphism from the group $G$ to
$\mathbb Z_2$, which labels unitarity of operators. Let $G_0$ be the subgroup of unitary operators, then $V$ splits into irreps of $G_0$:
\begin{align}
V=\bigoplus _{i} V_i\otimes \mathbb C^{m_i}
\end{align}
where $V_i$ are irreps and $m_i$ are their multiplicities in $V$. If there is no antiunitary operators then followed by Schur's lemma, the most general Hamiltonians are those belonging to the set
\begin{align}
\bigoplus _{i} \text{End}_G(V_i\otimes \mathbb C^{m_i})=\bigoplus _{i} \text{End}(\mathbb C^{m_i})
\end{align}
plus Hermicity. This is called Type A in the Altland-Zirnbauer theory, without any anti-unitary operators. The case with the presence of anti-unitary operators is more complicated. Let $T$ be an antiunitary operator, then the conjugation by $T$, i.e. $U\mapsto TUT^{-1}$, is an automorphism of $G_0$, thus $T$ maps a component $V_i\otimes \mathbb C^{m_i}$ to another $V_j\otimes \mathbb C^{m_j}$. A simple case is when $i\neq j$, which is easy to see that the most general Hamiltonian is of form \cite{Zirnbauer1996,AZ1997}
\begin{align}
(H,THT^{-1})
\end{align}
where $H$ is an Hermitian operator in component $i$ and $THT^{-1}$ acts on component $j$. Thus it's also of Type A.
\\
\\
Type A is the simplest structure without any further symmetries. However, if we consider $i=j$, and consider more anti-unitary operators, the situation is much more technical. It turns out that possible Hamiltonians with specific symmetric structures can be classified into ten classes. Here we skip the detailed analysis and directly present the final results. These classes are classified by the following three different operators,
\begin{itemize}
\item $T_+$, antiunitary, commutes with Hamiltonian, and $T_+^2=\pm 1$
\item $T_-$, antiunitary, anticommutes with Hamiltonian, and $T_-^2=\pm 1$
\item $\Lambda$, unitary, anticommutes with Hamiltonian, and $\Lambda^2=1$
\end{itemize}
If two of these three operators exist, the third will be determined by the following identity,
\begin{align}
\Lambda=T_+ T_-
\end{align}
The properties of these three operators can classify the Hamiltonian in the following ten classes,
\begin{center}
\begin{tabular}{ c | c | c | c | c | c }
$T_+^2$ &  $T_-^2$ & $\Lambda^2$ & Cartan label & Block & Type \\
\hline
 &  &  & A (GUE) & $M \text{  complex: } M^\dagger=M$ & $\mathbb{C}$  \\
1 &  &  & AI (GOE)  & $M \text{  real: } M^T=M$ & $\mathbb{R}$ \\
$-1$ &  &  & AII (GSE) & $M \text{  quaternion: } M^\dagger=M$ & $\mathbb{H}$ \\
 &  & 1 & AIII (chGUE) & $\left( \begin{matrix}
   0 & Z  \\
   Z^\dagger & 0  \\
\end{matrix} \right)
\text{   } Z \text{  complex } $ & $\mathbb{C}$ \\
 & $-1$ &  & C (BdG) & $\left( \begin{matrix}
   A & B  \\
   \bar{B} & -\bar{A}  \\
\end{matrix} \right) \begin{matrix}
   A \text{ Hermitian}  \\
   B \text{ complex symmetric}  \\
\end{matrix}
$ & $\mathbb{C}$  \\
 & 1 &  & D (BdG) & $M \text{  pure imaginary, skew-symmetric}$ & $\mathbb{C}$ \\
1 & 1 & 1 & BDI (chGOE) & $\left( \begin{matrix}
   0 & A  \\
   {{A}^{T}} & 0  \\
\end{matrix} \right)\text{   } A \text{  real }$ & $\mathbb{R}$ \\
1 & $-1$ & 1 & CI (BdG) & $\left( \begin{matrix}
   0 & Z  \\
   \bar{Z} & 0  \\
\end{matrix} \right)
\text{   } Z \text{  complex symmetric} $  & $\mathbb{R}$  \\
 $-1$ & 1 & 1 & DIII (BdG) & $\left( \begin{matrix}
   0 & Y  \\
   -\bar{Y} & 0  \\
\end{matrix} \right)\text{   } Y \text{  complex, skew-symmetric}$  & $\mathbb{H}$\\
 $-1$ & $-1$ & 1 & CII (chGSE) & $\left( \begin{matrix}
   0 & B  \\
   B^\dagger & 0  \\
\end{matrix} \right)\text{   } B \text{ quaternion}$  & $\mathbb{H}$\\
\end{tabular}
\end{center}
When there are no values in some corresponding operators, we mean that there is no such symmetry in the system. We also present the block representation in this table, where blocks are classified by the $\pm 1$ eigenspace of anti-unitary operators. The first three ensembles in this table are original Dyson ensembles, while other extended ensembles are their subsets with higher symmetries. These classifications are widely used in theoretical physics, for example, the symmetry classifications of topological insulators and topological phases \cite{ludwig,ki}.
\section{Eigenvalue distribution}\label{Dist}
This appendix is a simple introduction to the random matrix theory eigenvalue distribution (for instance, see \cite{Wishart,MIT}), the measure in the eigenvalue basis. For Wigner-Dyson ensemble, this is given by the formula
\begin{align}
P(\lambda)d\lambda=C(N,\tilde{\beta})|\Delta (\lambda)|^{\tilde{\beta}}\prod _{k}e^{-\frac{N\tilde{\beta}}{4}\lambda _k^2}d\lambda _k
\end{align}
where $\lambda =(\lambda _1,\cdots,\lambda _N)$ is the set of eigenvalues, $\Delta (\lambda)$ is the Vandermont determinant defined by
\begin{align}
\Delta (\lambda)=\prod _{k>l}(\lambda _k-\lambda _l)
\end{align}
and $C(N,\tilde{\beta})$ is a normalization constant depending on $\tilde{\beta}$ and $N$. For different ensembles, $\tilde{\beta}$ is defined as
\begin{center}
\begin{tabular}{ c | c   }
 RMT & $\tilde{\beta}$ \\
\hline
AI(GOE) & 1 \\
A(GUE) & 2 \\
AII(GSE) & 4 \\
\end{tabular}
\end{center}
For the remaining ensembles, the eigenvalues occur in pairs (because the $T_{-}$ operator introduced in the last appendix anti-commutes with $Q$), and the eigenvalues probability distribution is given by
\begin{align}\label{DistQ}
P(\lambda)d\lambda=C(N,\tilde{\beta},\tilde{\alpha})|\Delta (\lambda ^2)|^{\tilde{\beta}}\prod _{k}\lambda _k ^{\tilde{\alpha}}e^{-\frac{N\tilde{\beta}}{4}\lambda _k^2}d\lambda _k
\end{align}
where we only take the positive one from a pair of eigenvalues, and $C(N,\tilde{\beta},\tilde{\alpha})$ is defined also as the corresponding normalization constant. In the Altland-Zirnbauer classification, constants $\tilde{\alpha}$ and $\tilde{\beta}$ are set as (considering the real model of us, we have set the flavor number $N_f=0$ and the topological index $\nu=0$ in chiral ensembles)
\begin{center}
\begin{tabular}{ c | c | c  }
 RMT & $\tilde{\beta}$ & $\tilde{\alpha}$\\
\hline
BDI(chGOE) & 1 & 0\\
AIII(chGUE) & 2 & 1  \\
CII(chGSE) & 4 & 3 \\
CI(BdG) & 1 & 1\\
D(BdG) & 2 & 0 \\
C(BdG) & 2  & 2\\
DIII(BdG) & 4 & 1  \\
\end{tabular}
\end{center}
We will also need the eigenvalue distribution of the Hamiltonian which is the square of $Q$, so we can take the square distribution of \ref{DistQ}, which will change Gaussian distribution to Wishart-Laguerre, which is
\begin{align}\label{DistH}
P(\lambda)d\lambda=C'(N,\tilde{\beta},\tilde{\alpha})|\Delta (\lambda)|^{\tilde{\beta}}\prod _{k}\lambda _k ^{\frac{\tilde{\alpha}-1}{2}}e^{-\frac{N\tilde{\beta}}{4}\lambda _k}d\lambda _k
\end{align}
here $\lambda _k$ are nonnegative and $C'(N,\tilde{\beta},\tilde{\alpha})$ is a new normalization constant which is one half of $C(N,\tilde{\beta},\tilde{\alpha})$. We could also write
\begin{align}\label{DistH}
P(\lambda)d\lambda\sim |\Delta (\lambda)|^{\tilde{\beta}}\prod _{k}\lambda _k ^{\tilde{\mu}}e^{-\frac{N\tilde{\beta}}{4}\lambda _k}d\lambda _k
\end{align}
where $\tilde{\mu}=(\tilde{\alpha}-1)/2$. The following table summarize the related index for supersymmetric SYK model
\begin{center}
\begin{tabular}{ c | c | c | c | c}
$N\text{ mod } 8$ & $Q$ & $\tilde{\alpha}$ & $\tilde{\beta}$ & $\tilde{\mu}$ \\
\hline
0 & BDI (chGOE) & 0 & 1 & $-1/2$\\
2 & DIII (BdG) & 1 & 4 & $0$\\
4 & CII (chGSE) & 3 & 4 & $1$\\
6 & CI (BdG) & 1 & 1 & $0$
\end{tabular}
\end{center}
In $N \mod 8=0,4$, the index $\tilde{\mu}$ precisely matches Wishart matrix. Moreover, Although the result has $\tilde{\mu}$ dependence for $N \mod 8=2,6$, which does not precisely match Wishart matrix from Dyson Gaussian ensemble by index $\tilde{\mu}$, we could also use the terminology LOE/LSE to refer the universal class from squaring of Gaussian matrix, similar with Altland-Zirnbauer classification as a subset of Dyson, regardless multiple anti-unitary symmetries. Thus, we call $N \text{ mod } 8=0,2,4,6$ as LOE/LSE/LSE/LOE respectively,

\end{document}